\documentclass[a4paper,10pt,twoside]{cpc-hepnp}

\usepackage{CJK,upgreek,fancyhdr}
\usepackage{amssymb}
\usepackage{amsthm}
\usepackage[mathlines]{lineno}
\usepackage{graphicx}
\usepackage{units}
\usepackage{url}
\usepackage{amsmath}
\usepackage{amsfonts}
\usepackage{textcomp}
\usepackage{subfigure}
\usepackage{verbatim}
\usepackage{rotating}
\usepackage[colorlinks,linkcolor=blue,citecolor=blue]{hyperref}
\usepackage{float}
\usepackage{epsfig}
\usepackage{bm}
\usepackage{color}
\usepackage{pstricks}
\usepackage{pst-node}
\usepackage{times}
\usepackage{indentfirst}
\usepackage[english]{babel}
\usepackage[T1]{fontenc}
\usepackage{bm}

\usepackage{mathrsfs}
\usepackage{xspace}

\usepackage{multicol}

\newcommand{\dsonem}{D_{s1}(2536)^-}
\newcommand{\dstwom}{D^*_{s2}(2573)^-}
\newcommand{\dzerobar}{\overline{D}{}^{0}}
\newcommand{\dstzerobar}{\overline{D}{}^{*0}}
\newcommand{\dspp}{D^+_s}
\newcommand{\lamdcp}{\Lambda^+_c}
\newcommand{\lamdcm}{\overline{\Lambda}{}^-_c}
\newcommand{\gev}{\,\unit{GeV}}

\newcommand{\gevcc}{\,\unit{GeV}/c^2}
\newcommand{\mevcc}{\,\unit{MeV}/c^2}
\newcommand{\mev}{\,\unit{MeV}}

\newcommand{\pb}{\,\unit{pb}}
\newcommand{\dwave}{$D$-wave }
\newcommand{\swave}{$S$-wave }

\newcommand {\ie}{\emph{i}.\emph{e}.\xspace}

\begin{document}
\begin{CJK*}{UTF8}{gkai}

\fancyhead[c]{\small Chinese Physics C~~~Vol. **, No. * (20**)
******} \fancyfoot[C]{\small 010201-\thepage}

\footnotetext[0]{Received *** 20**}

\title{\boldmath Observation of $e^+e^- \rightarrow D_s^+ \overline{D}{}^{(*)0} K^-$ and study of the $P$-wave $D_s$ mesons}

\maketitle
\begin{small}
\begin{center}
M.~Ablikim(麦迪娜)$^{1}$, M.~N.~Achasov$^{10,d}$, S. ~Ahmed$^{15}$, M.~Albrecht$^{4}$, M.~Alekseev$^{55A,55C}$, A.~Amoroso$^{55A,55C}$, F.~F.~An(安芬芬)$^{1}$, Q.~An(安琪)$^{42,52}$, Y.~Bai(白羽)$^{41}$, O.~Bakina$^{27}$
% ...
\\
\vspace{0.2cm}
(BESIII Collaboration)\\
\vspace{0.2cm} {\it
$^{1}$ Institute of High Energy Physics, Beijing 100049, People's Republic of China\\
% ...
$^{4}$ Bochum Ruhr-University, D-44780 Bochum, Germany\\
% ...
$^{10}$ G.I. Budker Institute of Nuclear Physics SB RAS (BINP), Novosibirsk 630090, Russia\\
% ...
$^{15}$ Helmholtz Institute Mainz, Johann-Joachim-Becher-Weg 45, D-55099 Mainz, Germany\\
% ...
$^{27}$ Joint Institute for Nuclear Research, 141980 Dubna, Moscow region, Russia\\
% ...
$^{41}$ Southeast University, Nanjing 211100, People's Republic of China\\
$^{42}$ State Key Laboratory of Particle Detection and Electronics, Beijing 100049, Hefei 230026, People's Republic of China\\
% ...
$^{52}$ University of Science and Technology of China, Hefei 230026, People's Republic of China\\
% ...
$^{55}$ (A)University of Turin, I-10125, Turin, Italy; (B)University of Eastern Piedmont, I-15121, Alessandria, Italy; (C)INFN, I-10125, Turin, Italy\\
% ...
$^{d}$ Also at the Novosibirsk State University, Novosibirsk, 630090, Russia\\
% ...
}\end{center}

\vspace{0.4cm}
\end{small}

\begin{abstract}
Studies of $e^+e^- \to \dspp \overline{D}{}^{(*)0}K^-$ and the
$P$-wave charmed-strange mesons are performed based on an $e^+e^-$
collision data sample corresponding to an integrated luminosity of 567
pb$^{-1}$ collected with the BESIII detector at $\sqrt{s}= 4.600 \gev$.
The processes of $e^+e^-\to\dspp\dstzerobar K^-$ and $\dspp\overline{D}{}^{0} K^-$ are observed for the first time and are found to be dominated by the modes $D_s^+\dsonem$ and $D_s^+\dstwom$, respectively.
The Born cross sections are measured to be $\sigma^{B}(e^+e^-\to\dspp\dstzerobar K^-) = (10.1\pm2.3\pm0.8)~\rm\pb$ and $\sigma^{B}(e^+e^-\to\dspp \overline{D}{}^{0} K^-) = (19.4\pm2.3\pm1.6)~\rm\pb$, and the products of Born cross section and the decay branching fraction are measured to be
$\sigma^{B}(e^+e^-\to\dspp\dsonem + c.c.)\cdot\mathcal{B}(\dsonem\to\dstzerobar K^-) = (7.5 \pm 1.8 \pm 0.7)~\rm\pb$
and
$\sigma^{B}(e^+e^-\to\dspp\dstwom + c.c.)\cdot\mathcal{B}(\dstwom\to\dzerobar K^-) = (19.7 \pm 2.9 \pm 2.0)~\rm\pb$.
For the $\dsonem$ and $\dstwom$ mesons, the masses and widths are measured to be
$M(\dsonem) = (2537.7 \pm 0.5 \pm 3.1)~\mevcc,$
$ \Gamma(\dsonem) = (1.7\pm 1.2 \pm 0.6)~\rm\mev,$
and
$M(\dstwom) = (2570.7\pm 2.0 \pm 1.7)~\mevcc,$ $\Gamma(\dstwom) = (17.2 \pm 3.6 \pm 1.1)~\rm\mev.$
The spin-parity of the $\dstwom$ meson is determined to be $J^P=2^{+}$.
In addition, the process $e^+e^-\to \dspp \overline{D}{}^{(*)0} K^-$ are searched for using the data samples taken at four (two) center-of-mass energies between 4.416 (4.527) and 4.575 GeV, and upper limits at the $90\%$ confidence level on the cross sections are determined.
\end{abstract}

\begin{keyword}
cross section, $P$-wave $D_s$ mesons, resonance parameters, spin-parity,  BESIII
\end{keyword}

\begin{pacs}
 14.40 Lb, 13.66 Bc
 \end{pacs}

\begin{multicols}{2}


\section{Introduction}
Although the Heavy Quark Effective Theory (HQET)~\cite{model1, model2, model3, model4} has achieved great success in the past decades in explaining and predicting the spectrum of charmed-strange mesons ($D_s$), there still exist discrepancies between the theoretical predictions and experimental measurements, especially for the $P$-wave excited states.
The unexpectedly low masses of $D_{s0}^{*}(2317)^-$ and $D_{s1}(2460)^-$ stimulated theoretical and experimental interest not only in them, but also in the other two $P$-wave charmed-strange states, $D_{s1}(2536)^-$ and $D_{s2}(2573)^-$.
The resonance parameters of the $\dsonem$ and $\dstwom$ mesons need more experimentally independent measurements~\cite{pdg}.
In particular, the latest result on the $\dstwom$ mass from LHCb~\cite{lhcds2mw,Aaij:2014xza} deviates from the other measurements~\cite{pdgDiff1, pdgDiff2, pdgDiff3} significantly, and therefore, the world average fit gives a bad quality $\chi^2/ndf=17.1/4$~\cite{pdg}, where $ndf$ is the number of degrees of freedom.
In addition, the quantum numbers spin and parity ($J^P$) of the $\dstwom$ meson have been determined to be $J^P=2^+$ only recently with a partial wave analysis carried out by LHCb~\cite{lhcbds2jp},  and more confirmation is needed.

In recent years, measurements of the exclusive cross sections for $e^+e^-$ annihilation into charmed or charmed-strange mesons above the open charm threshold have attracted great interest.
First, the charmonium states above the open charm threshold ($\psi$ states) still lack of adequate experimental measurements and theoretical explanations.
The latest parameter values of these $\psi$ resonances are given by BES~\cite{besRvalue} from a fit to the total cross section of hadron production in $e^+e^-$ annihilation.
However, model predictions for $\psi$ decays into two-body final states were used, hence the values of the resonance parameters remain model-dependent.
Studies of the exclusive $e^+e^-$ cross sections would help to measure the parameters of the $\psi$ states model-independently.
Second, many additional $Y$ states with $J^P = 1^{--}$ lying above the open charm threshold have been discovered recently~\cite{Y1, Y2, Y3, Y4, Y5}.
Exclusive cross section measurements will provide important information in explaining these states.
Measurements of $e^+e^-$ cross sections for the $D_{(s)}^{(*)}\overline{D}{}^{(*)}_{(s)}$ final states were performed by Belle~\cite{dd1, dd2, dd3, dd4,dd5, dd6}, \textit{BABAR}~\cite{dd7, dd8, dd9}, and CLEO~\cite{dd10}, only with low-lying charmed or charmed-strange mesons in the final states.
Up to now, only the $D\overline{D}{}^{*}_{2}(2460)$ final states in $e^+e^-$ annihilation have been observed by Belle~\cite{d2d}, others with higher excited charmed  or charmed-strange mesons have not yet been observed.
In addition, the cross sections of $e^+e^-\to D\overline{D}{}^{(*)}\pi$ have also been measured by CLEO~\cite{dd10} and BESIII~\cite{dd11, dd12, dd13, dd14}.
However, a search for final states with strange flavor, $e^+e^-\to D_{s}^+ \overline{D}{}^{(*)0}K^-$, has not been performed before.

Using $e^+e^-$ collision data corresponding to an integrated luminosity of 567 $\rm pb^{-1}$~\cite{lum} collected  at a center-of-mass energy of $\sqrt{s}=4.600$ GeV with the BESIII detector operating at the Beijing Electron-Positron Collider (BEPCII), we observe the processes $e^+e^- \to \dspp \overline{D}{}^{*0}K^-$ and $e^+e^- \to \dspp \overline{D}{}^{0}K^-$, which are found to be dominated by $D_s^+ \dsonem$ and $D_s^+ \dstwom$, respectively.
For the observed $\dsonem$ and $\dstwom$ mesons, we present the resonance parameters and determine the spin and parity of $\dstwom$.
In addition, the processes $e^+e^-\to \dspp \overline{D}{}^{(*)0} K^-$ are searched for using the data samples taken at four (two) center-of-mass energies between 4.416 (4.527) and 4.575 GeV, and upper limits at $90\%$ confidence level on the cross sections are determined.
Throughout the paper, the charge conjugate processes are implied to be included, unless explicitly stated otherwise.

\section{BESIII Detector and Monte Carlo Simulation}
\label{sec:besiii}

The BESIII detector is a magnetic spectrometer~\cite{Ablikim:2009aa} located at the Beijing Electron Positron Collider (BEPCII)~\cite{Yu:IPAC2016-TUYA01}. The cylindrical core of the BESIII detector consists of a helium-based multilayer drift chamber (MDC), a plastic scintillator time-of-flight  system (TOF), and a CsI(Tl) electromagnetic calorimeter (EMC), which are all enclosed in a superconducting solenoidal magnet providing a 1.0~T magnetic field.
The solenoid is supported by an octagonal flux-return yoke with resistive plate counter muon identifier modules interleaved with steel. The acceptance for charged particles and photons is 93\% over $4\pi$ solid angle. The charged-particle momentum resolution at $1~{\rm GeV}/c$ is $0.5\%$, and the specific energy loss ($dE/dx$) resolution is $6\%$ for electrons from Bhabha scattering. The EMC measures photon energies with a resolution of $2.5\%$ ($5\%$) at $1$~GeV in the barrel (end cap) region. The time resolution of the TOF barrel part is 68~ps, while that of the end cap part is 110~ps.

Simulated data samples are produced with the {\sc geant4}-based~\cite{geant4} Monte Carlo (MC) package which includes the geometric description of the BESIII detector and the detector response.  They are used to determine the detection efficiency and to estimate the backgrounds. The simulation includes the beam energy spread and effects of initial state radiation (ISR) in the $e^+e^-$ annihilations modeled with the generator {\sc kkmc}~\cite{ref:kkmc}.
The inclusive MC samples consist of the production of open charm processes, the ISR production of vector charmonium(-like) states, and the continuum processes incorporated in {\sc kkmc}~\cite{ref:kkmc}.
The known decay modes are model-led with {\sc evtgen}~\cite{ref:evtgen} using branching fractions taken from the Particle Data Group~\cite{pdg}, and the remaining unknown decays from the charmonium states with {\sc lundcharm}~\cite{ref:lundcharm}. Final state radiation (FSR) from charged final state particles is simulated with the {\sc photos} package~\cite{photos}.
The intermediate states in the $D_s^+\to K^+K^-\pi^+$ decay are considered in the simulation~\cite{dsdalitz}.
In the measurements of $\dsonem$ and $\dstwom$ resonance parameters, the angular distributions are taken into account in the generation of signal MC samples.
For the signal process of $e^+e^-\to D^+_s \dsonem, \dsonem \to \dstzerobar K^-$, the spin-parity of the $\dsonem$ meson is assumed to be $1^+$.
To determine the spin-parity of $\dstwom$, efficiencies were obtained from the two MC samples, which assume the spin-parity as $1^-$ or $2^+$.
The MC sample with spin-parity $2^+$ is used in the measurement of the $\dstwom$ resonance parameters.

\section{Basic event selections}
\label{sec:eventselection}
To identify the final state $D_s^+ \overline{D}{}^{(*)0} K^-$, a partial reconstruction method is adopted, in which  we  detect the $K^-$ and reconstruct $\dspp$ candidates through the $\dspp\to K^+ K^- \pi^+$ decay.
The remaining $\overline{D}{}^{(*)0}$ meson is identified with the mass recoiling against the reconstructed $K^-D_s^+$ system.

For each of the four reconstructed charged tracks, the polar angle in the MDC must satisfy $|\cos\theta|<0.93$, and the distance of the closest approach from the $e^+e^-$ interaction point to the reconstructed track is required to be within $10$ cm in the beam direction and within $1$ cm in the plane perpendicular to the beam direction.
The ionization energy loss $dE/dx$ measured in the MDC and the time of flight measured by the TOF are used to perform the particle identification (PID).
Pion candidates are required to satisfy
 $\rm{prob}(\pi)>\rm{prob}(K)$, where $\rm{prob}(\pi)$ and $\rm{prob}(K)$ are the PID confidence levels for a track to be a pion and kaon, respectively.
Kaon candidates are identified by requiring  $\rm{prob}(K)>\rm{prob}(\pi)$.

\begin{figure*}[tp]
\centering
\includegraphics[width=0.49\linewidth]{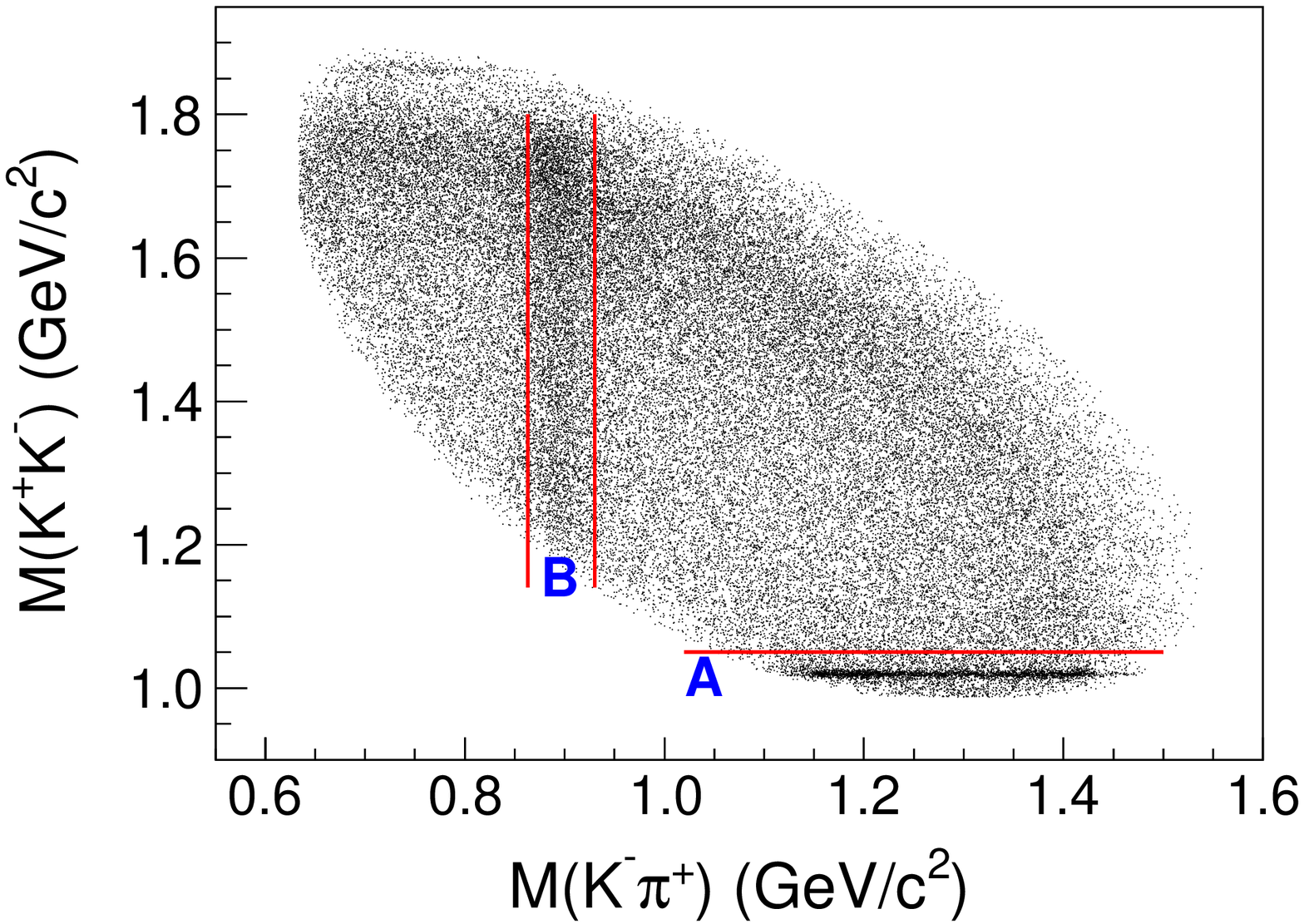}
	\put(-40,145){\bf (a)}
\includegraphics[width=0.49\linewidth]{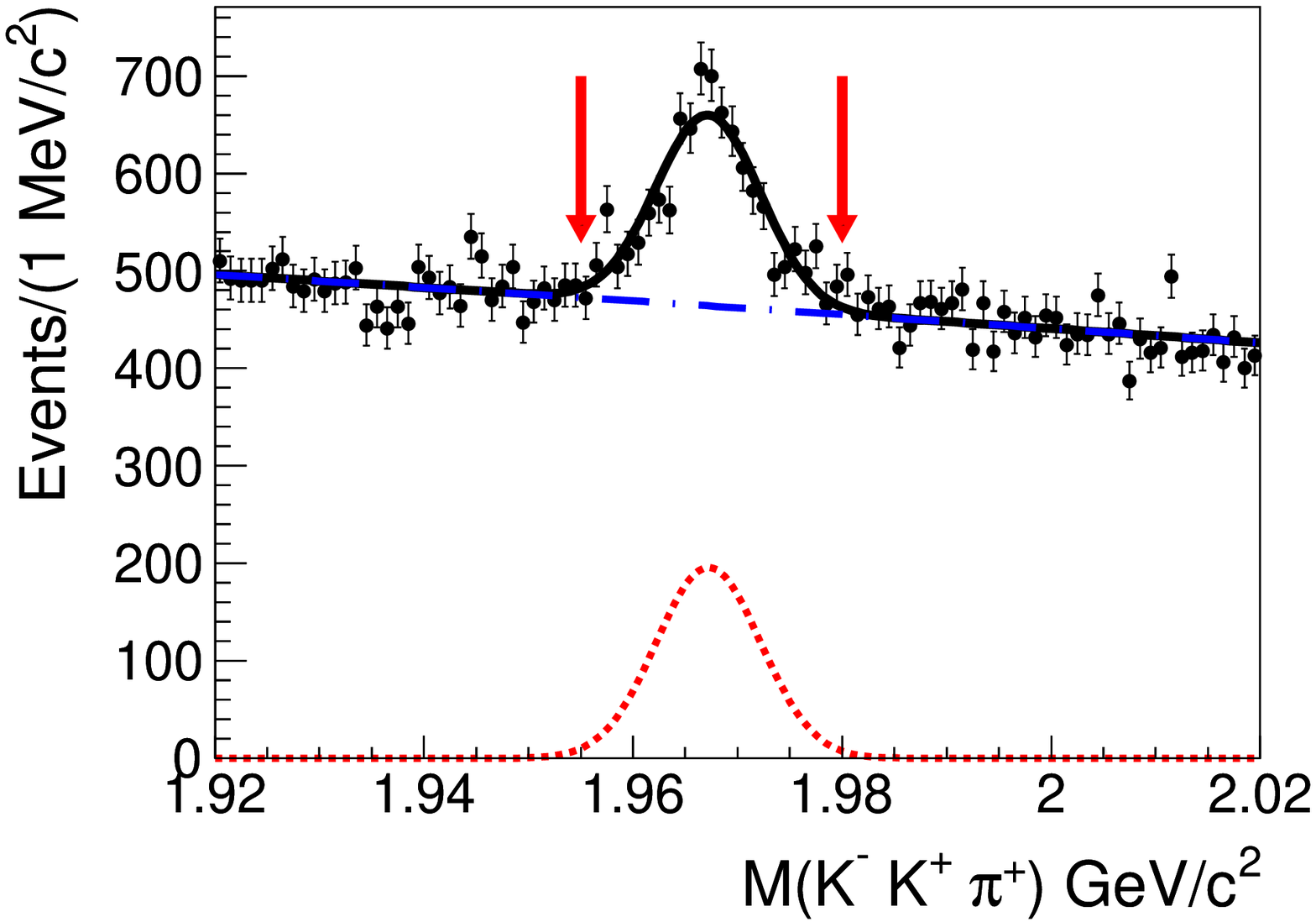}
	\put(-40,145){\bf (b)}\\
\includegraphics[width=0.49\linewidth]{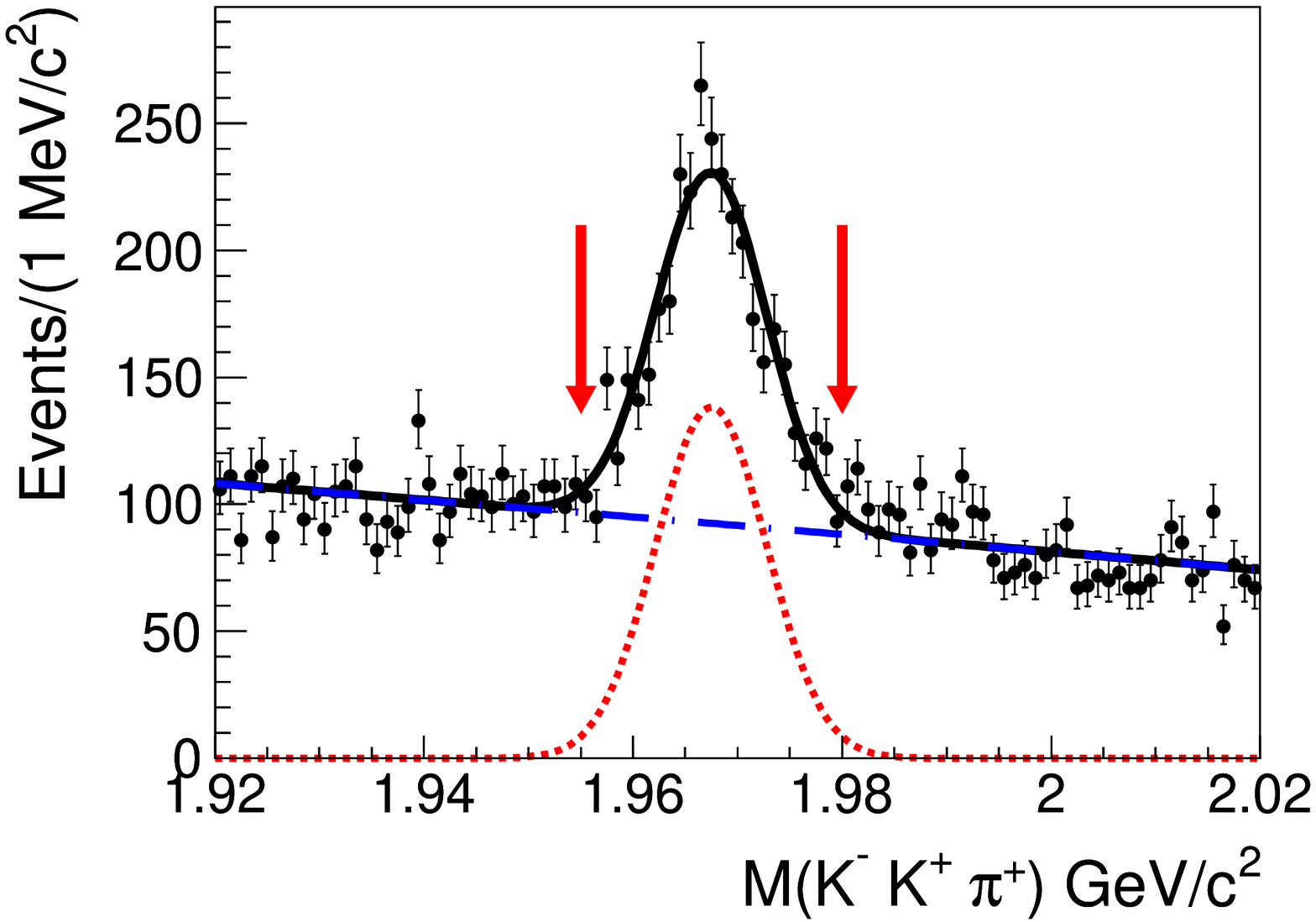}
	\put(-40,145){\bf (c)}
	\put(-65,135){regions A+B}
\includegraphics[width=0.49\linewidth]{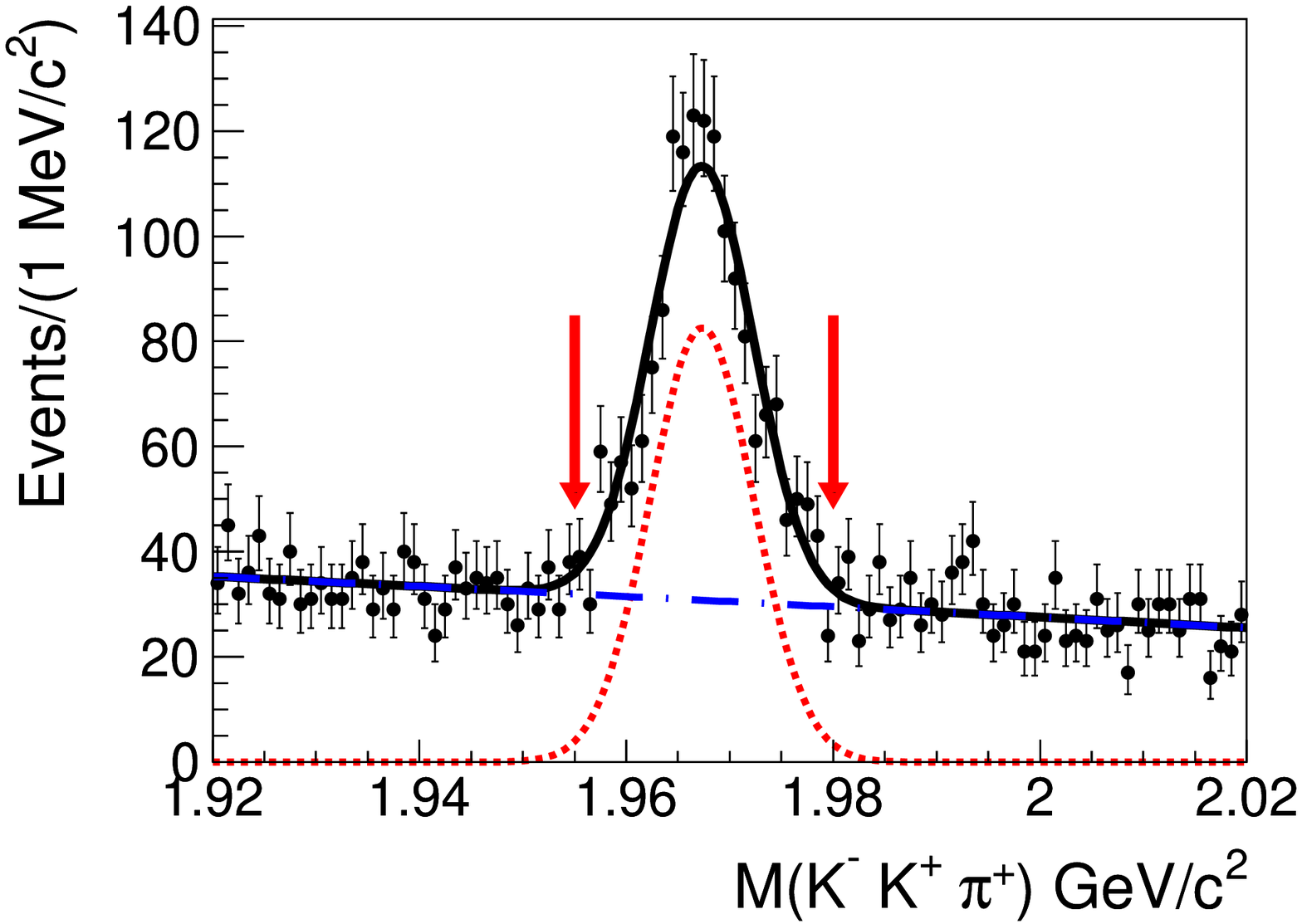}
	\put(-40,145){\bf (d)}
	\put(-55,135){region A}
\caption{
	Scatter plot of $M(K^+K^-)$ versus $M(K^-\pi^+)$ for the $D_s^+ \to K^+ K^- \pi^+$ candidates (a)
	and the corresponding invariant mass $M(K^+K^-\pi^+)$ distribution (b) for data at $\sqrt{s} = 4.600 \gev$. The $M(K^+K^-\pi^+)$ distributions of the subsamples from the regions A+B and  from the region A are shown in plot (c) and (d), respectively.
In plots (b), (c) and (d), fits with the sum of a Gaussian function and a  polynomial function are implemented to determine the signal regions for the $D_s^+$ candidates. The signal windows are shown with arrows.
}
\label{fig:dsscatterplot}
\end{figure*}

The $D_s^+$ meson candidates are reconstructed from two kaons with opposite charge and one charged pion.
To satisfy strangeness and charge conservation, each $D_s^+$ candidate must be accompanied by a negatively charged kaon.
For the $D_s^+$ candidates, the distributions of the reconstructed masses $M(K^+K^-)$ versus $M(K^-\pi^+)$ and $M(K^-K^+\pi^+)$ are shown in Figs.~\ref{fig:dsscatterplot}(a) and (b), respectively.
The two dominant sub-resonant decays, $\ie$, a horizontal band for the process $D_s^+ \to \phi \pi^+$ and a vertical band for the process $D_s^+ \to K^+ \overline{K}{}^{*}(892)^0$ are clearly visible.
To improve the signal significance in Fig.~\ref{fig:dsscatterplot}(b), only the $D_s^+$ candidates which satisfy $M(K^+K^-)<1.05~\gevcc$ (region A) or  $0.863 < M(K^-\pi^+) < 0.930~\gevcc$ (region B) are retained. The corresponding $M(K^-K^+\pi^+)$ distributions for events in region A+B and A are plotted in Figs.~\ref{fig:dsscatterplot}(c) and (d), respectively, showing improved signal significance.
The final $D_s^+$ candidates must have a reconstructed mass  $M(K^-K^+\pi^+)$ in the region $(1.955, 1.980)~\gevcc$.

In this analysis, the resolution of the recoiling mass is improved by using the variables
$RQ(K^-D_s^+)\equiv RM(K^-D_s^+) + M(D_s^+) - m(D_s^+)$ and $RQ(D_s^+) \equiv RM(D_s^+) + M(D_s^+) - m(D_s^+)$.
Here, $RM(D_s^+)$ and $RM(K^-D_s^+)$  are the reconstructed recoiling masses against the $D_s^+$ and $K^-D_s^+$ system, respectively, and  $m(D_s^+)$ is the nominal $D_s^+$ mass taken from the world average~\cite{pdg}.

\section{Studies of data at 4.600 GeV}

\subsection{Cross section of \boldmath $e^+e^-\to D_s^+ \overline{D}{}^{(*)0} K^-$ }
\label{sec:xs4600}
\begin{figure*}[tp]
	\centering
	\includegraphics[width=0.48\linewidth]{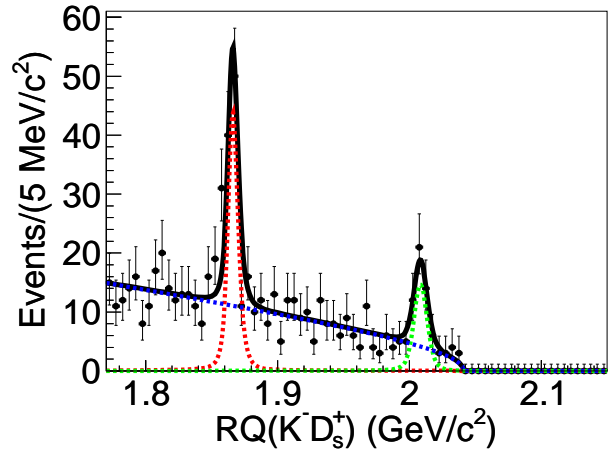}
	\caption{
	Distributions of $RQ(K^- D_s^+)$ for the $D_s^+$ signal candidates in regions A + B in Fig.~\ref{fig:dsscatterplot}(c), for data taken at $\sqrt{s} = 4.600\gev$. The solid line shows the total fit to the data points and the dashed lines represent the $\overline{D}{}^{0}$ and $\overline{D}{}^{*0}$ signals.
	}
	\label{fig:recoil_kds}
\end{figure*}

To reject the backgrounds from $\Lambda_c^{+}$ decays in the measurement of the cross section of $e^+e^-\to D_s^+ \overline{D}{}^{(*)0} K^-$, we further demand that $RQ(D_s^+)<2.59 \gevcc$.
Figure~\ref{fig:recoil_kds}  presents evident peaks in the distribution of $RQ(K^-D_s^+)$ around the signal positions of $\overline{D}{}^{*0}$ and $\overline{D}{}^{0}$, which correspond to the processes $e^+e^- \to \dspp \overline{D}{}^{*0}K^-$  and $\dspp \overline{D}{}^{0}K^-$, respectively.

To determine the signal yields of the processes $e^+e^-\to D_s^+ \overline{D}{}^{(*)0} K^-$ at 4.600 GeV, an unbinned maximum likelihood fit is performed to the $RQ(K^-D_s^+)$ spectrum as shown in Fig.~\ref{fig:recoil_kds}. The signal peaks are described by the MC-determined signal shapes and the background shapes are taken as ARGUS functions~\cite{argus}.
In the fit to data, the endpoint of the background shape is fixed at the value obtained from a fit of an ARGUS function to the $RQ(K^-D_s^+)$ spectrum in the background MC sample.
The Born cross section is calculated as
\begin{equation}
\sigma^{B} = \frac{N_{\rm obs}}{\mathcal{L}(1+\delta)\frac{1}{|1-\Pi|^2}\mathcal{B}\epsilon},
\end{equation}
where $N_{\rm obs}$ is the number of the observed signal candidates,
$\mathcal{L}$ is the integrated luminosity,
$\epsilon$ is the detection efficiency determined from MC simulations,
$(1+\delta)$ is the radiative correction factor~\cite{Kuraev:1985hb},
$\frac{1}{|1-\Pi|^2}$ is the vacuum polarization factor~\cite{vacuum.polarization},
and $\mathcal{B}$ is branching fraction of $\dspp\to K^+ K^- \pi^+$.
The detection efficiencies are estimated based on MC simulations, assuming the two body final states of $ D_s^+\dsonem$ and $D_s^+\dstwom$ dominate the decays to $D_s^+ \overline{D}{}^{(*)0} K^-$ according to the studies in Secs.~\ref{sec:dsone} and ~\ref{sec:dstwo}. The numerical results are given in Table~\ref{table:3bodycrosssection}.

\subsection{\boldmath Studies on the $\dsonem$}
\label{sec:dsone}

\begin{figure*}[tp]
	\centering
	\includegraphics[width=0.45\linewidth]{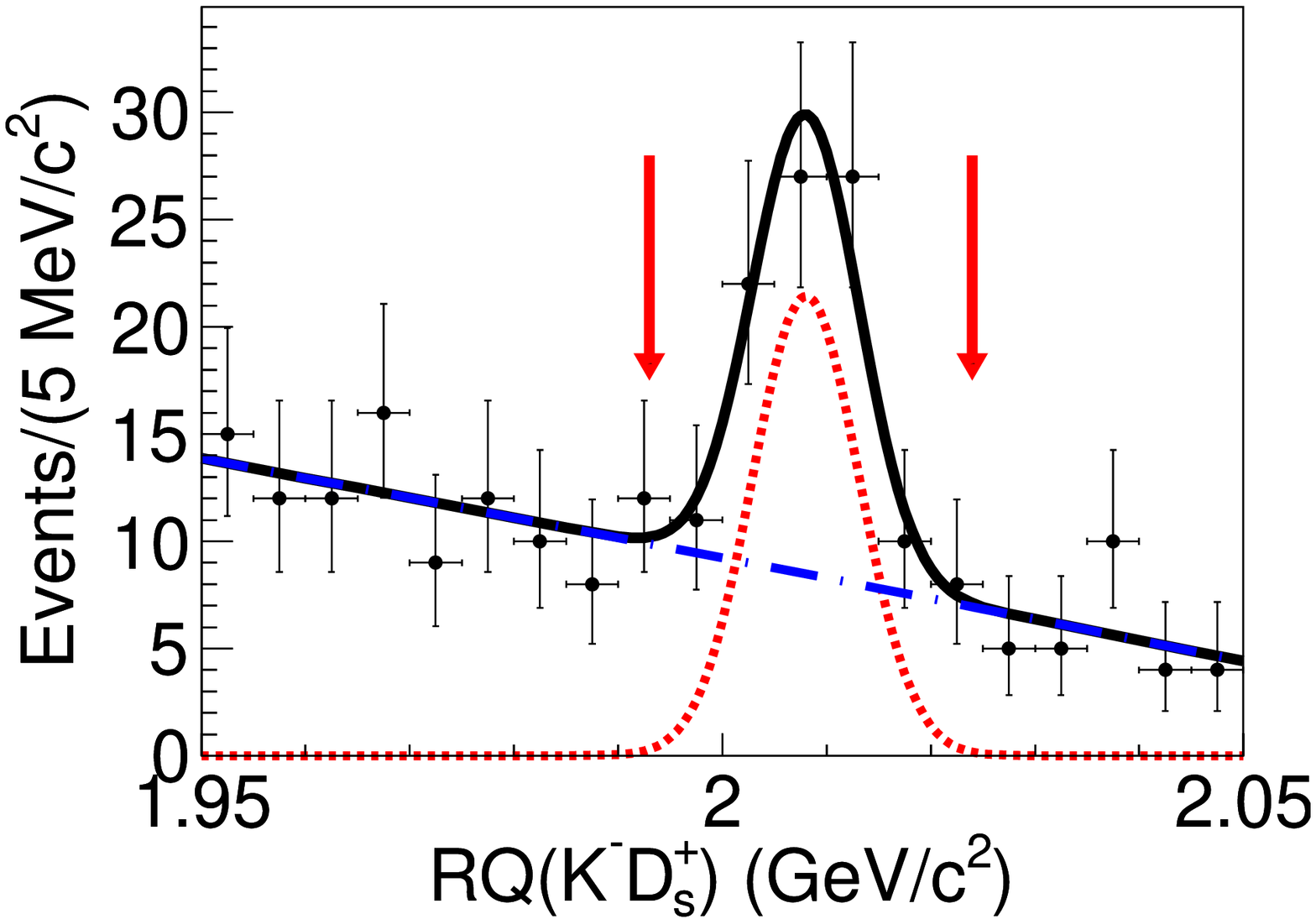}
	\put(-40,140){\bf (a)}
	\includegraphics[width=0.45\linewidth]{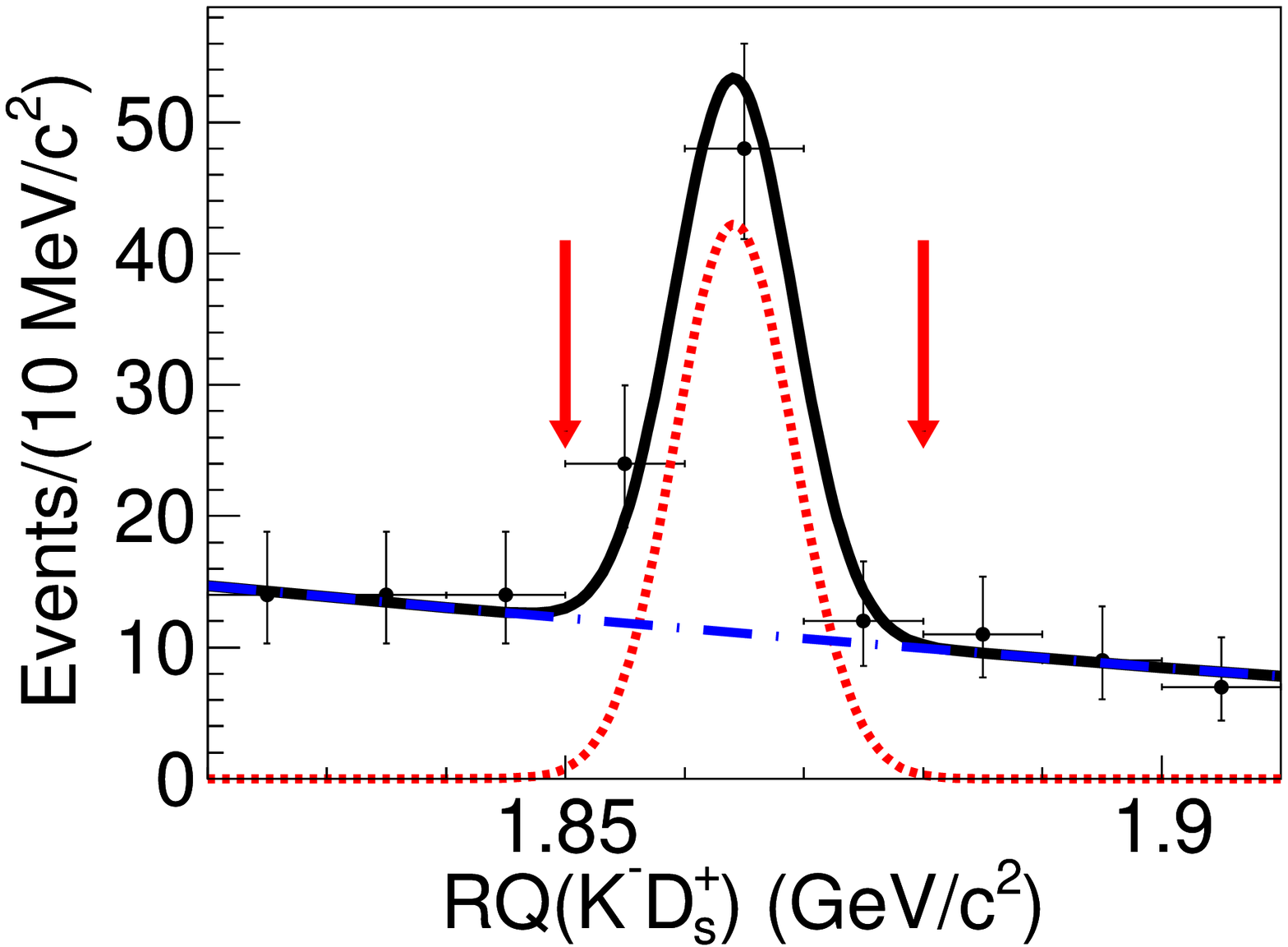}
	\put(-40,140){\bf (b)}
	\caption{
	At 4.600 GeV, (a) the $RQ(K^- D_s^+)$ distribution for the $D_s^+$ candidates from signal regions A and B in Fig.~\ref{fig:dsscatterplot}(c); (b) the $RQ(K^- D_s^+)$ distribution for the $D_s^+$ candidates from signal regions A in Fig.~\ref{fig:dsscatterplot}(d).  Fits with the sum of a Gaussian function and a polynomial function are implemented to determine the signal regions for the $\overline{D}{}^{(*)0}$ candidates, which are indicated with arrows.}
	\label{fig:dsksel}
\end{figure*}

\begin{figure*}[tp]
	\centering
	\includegraphics[width=0.48\linewidth]{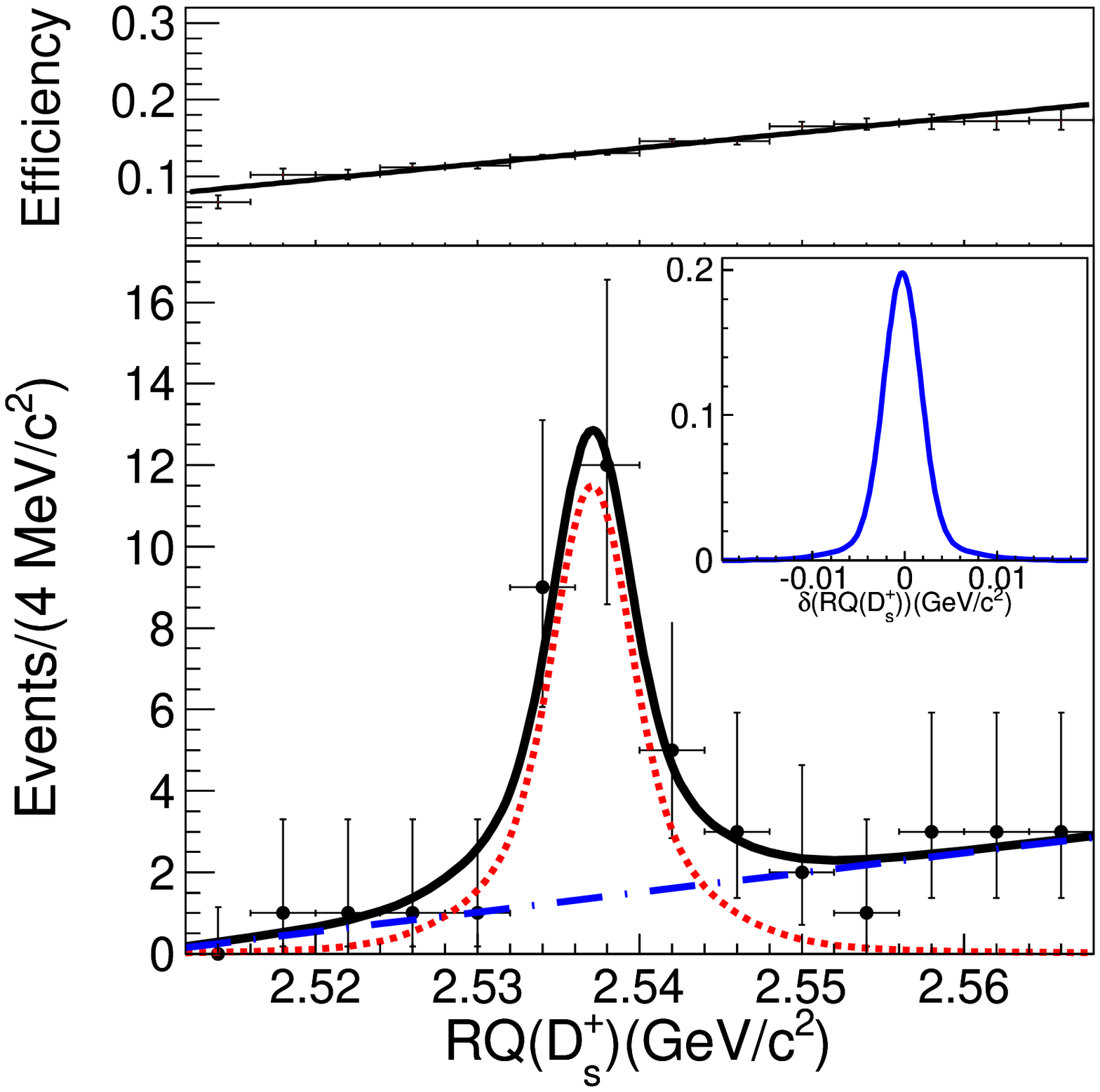}
	\put(-44, 151){\bf (c)}
	\put(-44, 185){\bf (b)}
	\put(-170,151){\bf (a)}
	\includegraphics[width=0.48\linewidth]{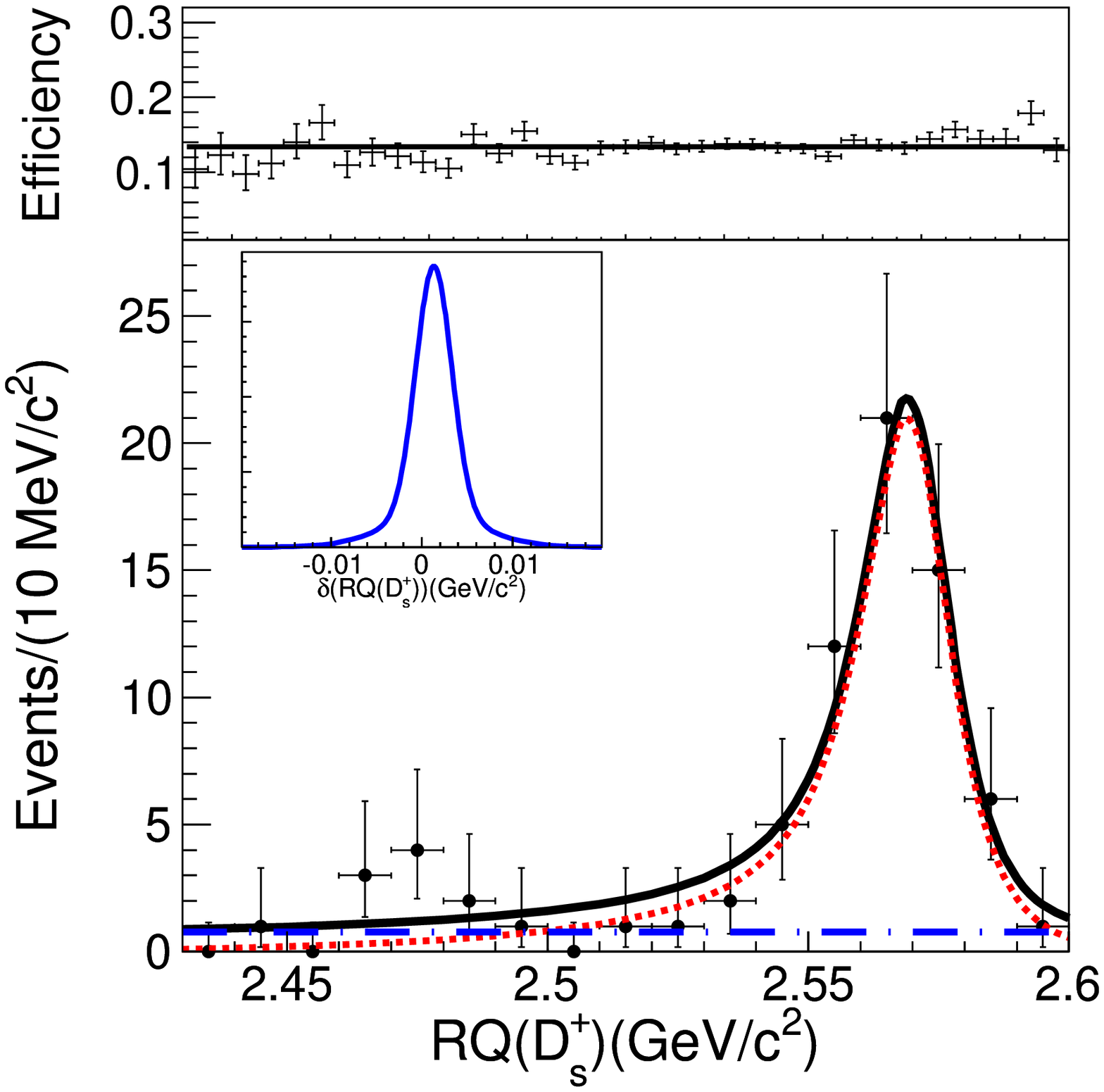}
	\put(-44, 151){\bf (d)}
	\put(-44, 185){\bf (e)}
	\put(-170,151){\bf (f)}
	\caption{
	At 4.600 GeV, the $RQ(D_s^+)$ spectra in the samples of $e^+e^-\to D_s^+ \overline{D}^{*0} K^-$ (left) and $e^+e^-\to D_s^+ \overline{D}^{0} K^-$ (right). Plots (a) and (d) show the result of the unbinned maximum likelihood fits.  Data are denoted by the dots with error bars. The dash-dotted and dotted lines are the background and signal contributions, respectively. Plots (b) and (e) show the efficiency functions. Plots (c) and (f) show the $RQ(D_s^+)$ resolution functions determined from MC simulations. }
	\label{fig:dsfit}
\end{figure*}

For the candidates surviving the basic event selections, we further select the signal candidates for $e^+e^-\to D_s^+ \overline{D}{}^{*0} K^-$  by requiring $1.993 < RQ(K^- D_s^+) < 2.024~\gevcc$, as shown in Fig.~\ref{fig:dsksel}(a).
The $RQ(D_s^+)$ distribution of the remaining events is displayed in Fig.~\ref{fig:dsfit}(a), where a clear $\dsonem$ signal peak near the nominal $\dsonem$ mass is visible.
An unbinned maximum likelihood fit is performed to the distribution, where
the signal shape is taken as a sum of the efficiency-weighted \dwave and \swave Breit-Wigner function convolved with the detector resolution function,
$ [\mathscr{E} \cdot (f \cdot BW_{S} + (1-f)\cdot BW_{D} )] \otimes \mathscr{R} $.
Here, the resolution function $\mathscr{R}$ (plotted in Fig.~\ref{fig:dsfit}(c)) and the efficiency $\mathscr{E}$ (plotted in Fig.~\ref{fig:dsfit}(b) ) are determined from MC simulations, and $f$ is the fraction of the $S$-wave Breit-Wigner function.
 The $S$-wave  and $D$-wave Breit-Wigner functions are $BW_{S} = \frac{1}{(RQ^2 - m^2)^2 + m^2 \Gamma^2}  \cdot p \cdot q$, and $BW_{D} = \frac{1}{(RQ^2 - m^2)^2 + m^2 \Gamma^2} \cdot p^5 \cdot q$, respectively, where $m$ and $\Gamma$ are the mass and width of the $\dsonem$ to be determined and $p(q)$ is the momentum of $K^-$($D_s^+$) in the rest frame of $K^- \overline{D}{}^{*0}$($e^+e^-$) system.
The backgrounds are described with a first-order polynomial function.
The parameter $f$ is fixed to 0.72~\cite{Ds1mixing}, while the other parameters are determined in the fit.

In this fit, the number of signal candidates is estimated to be $24.0 \pm 5.7(\rm{stat})$.
The mass and width of the $\dsonem$ are measured to be $(2537.7 \pm 0.5(\rm{stat}) \pm 3.1(\rm{syst}))~\mevcc$, and $(1.7 \pm 1.2(\rm{stat}) \pm 0.6(\rm{syst}))~\mev$, respectively.
The branching fraction weighted Born cross section is determined to be $\sigma^{B}(e^+e^-\to\dspp\dsonem + c.c.)\cdot\mathcal{B}(\dsonem\to\dstzerobar K^-) = (7.5 \pm 1.8 \pm 0.7)~\rm\pb$.
The relevant systematic uncertainties are discussed later and summarized in Table~\ref{table:3bodycrosssectionErr}.

\subsection{\boldmath Studies on the $\dstwom$}
\label{sec:dstwo}

To study the $\dstwom$ properties, we select the signal candidates of the process $e^+e^-\to D_s^+ \overline{D}{}^{0} K^-$  by requiring  $RQ(K^- D_s^+)$ in the $\overline{D}{}^{0}$ signal region of $(1.850, 1.880)~\gevcc$, as shown in Fig.~\ref{fig:dsksel}(b).
To reject backgrounds from $e^+e^- \to \lamdcp \lamdcm$, only the $D_s^+$ candidates in region A of Fig.~\ref{fig:dsscatterplot} are used.
For the selected events, the corresponding $RQ(D_s^+)$ distribution is plotted in Fig.~\ref{fig:dsfit}(d), where a clear $\dstwom$ signal peak near the known $\dstwom$ mass is observed.

An unbinned maximum likelihood fit is performed to the $RQ(D_s^+)$ spectrum  in Fig.~\ref{fig:dsfit}(d).
The spin-parity of the $\dstwom$ meson is fixed to be $2^+$, following the studies in Sec.~\ref{sec:jpdstwo}, and the $\dstwom$ meson is assumed to decay to $\dzerobar K^-$ predominantly via $D$-wave~\cite{model2}.
Hence, we take the $D$-wave Breit-Wigner function $BW = \frac{1}{(RQ^2 - m^2)^2 + m^2 \Gamma^2} \cdot p^5 \cdot q^5$ convolved with the resolution function (shown in Fig.~\ref{fig:dsfit}(f)), $BW \otimes \mathscr{R}$, to describe the signal, and a flat line to represent backgrounds.
Here, $p(q)$ is the momentum of $K^-$($D_s^+$) in the rest frame of the $K^- \overline{D}{}^{0}$($e^+e^-$) system.
Figure~\ref{fig:dsfit} (e) shows the efficiency distribution with the assignment $J^P = 2^+$, which is consistent with a flat line.
All parameters are left free in the fit.

The fit yields $61.9 \pm 9.1(\rm{stat})$ signal events.
The mass and width of the $\dstwom$ are measured to be $(2570.7\pm  2.0(\rm{stat}) \pm 1.7(\rm{syst}))~\mevcc$, and $(17.2 \pm 3.6(\rm{stat}) \pm 1.1(\rm{syst}))~\mev$, respectively, where the systematic uncertainties are summarized in Table~\ref{table:ds1ds2uncertainty}.
The branching fraction weighted Born cross section is given to be
$\sigma^{B}(e^+e^-\to\dspp\dstwom + c.c.)\cdot\mathcal{B}(\dstwom\to\dzerobar K^-) = (19.7 \pm 2.9 \pm 2.0)~\rm\pb$.
The relevant systematic uncertainties are discussed later and summarized in Table~\ref{table:3bodycrosssectionErr}.

\subsection{\boldmath Spin-parity of the $\dstwom$}
\label{sec:jpdstwo}

At $\sqrt{s}= 4.600 \gev$, the exclusive process $e^+e^-\to\dspp\dstwom \to\dspp\dzerobar K^-$ is observed just above the production threshold.
For the $\dstwom$ meson, the $J^P$ assignments with high spins would be strongly suppressed in this process.
Hence, we assume that the $\dstwom$ meson can only have two possible $J^{P}$ assignments, $1^{-}$ or $2^{+}$.
Under these two hypotheses, the differential decay rates as a function of the helicity angle $\theta'$ of the $K^-$ in the rest frame of the $\dstwom$, $\rm dN / \rm d\cos\theta'$, follow two very distinctive formulae of $(1-\cos^2\theta')$ for $1^-$ and $\cos^2\theta'(1-\cos^2\theta')$ for $2^+$.
We can determine the true spin-parity from tests of the two hypotheses based on data.

\begin{figure*}
\centering
	\includegraphics[width=0.48\linewidth]{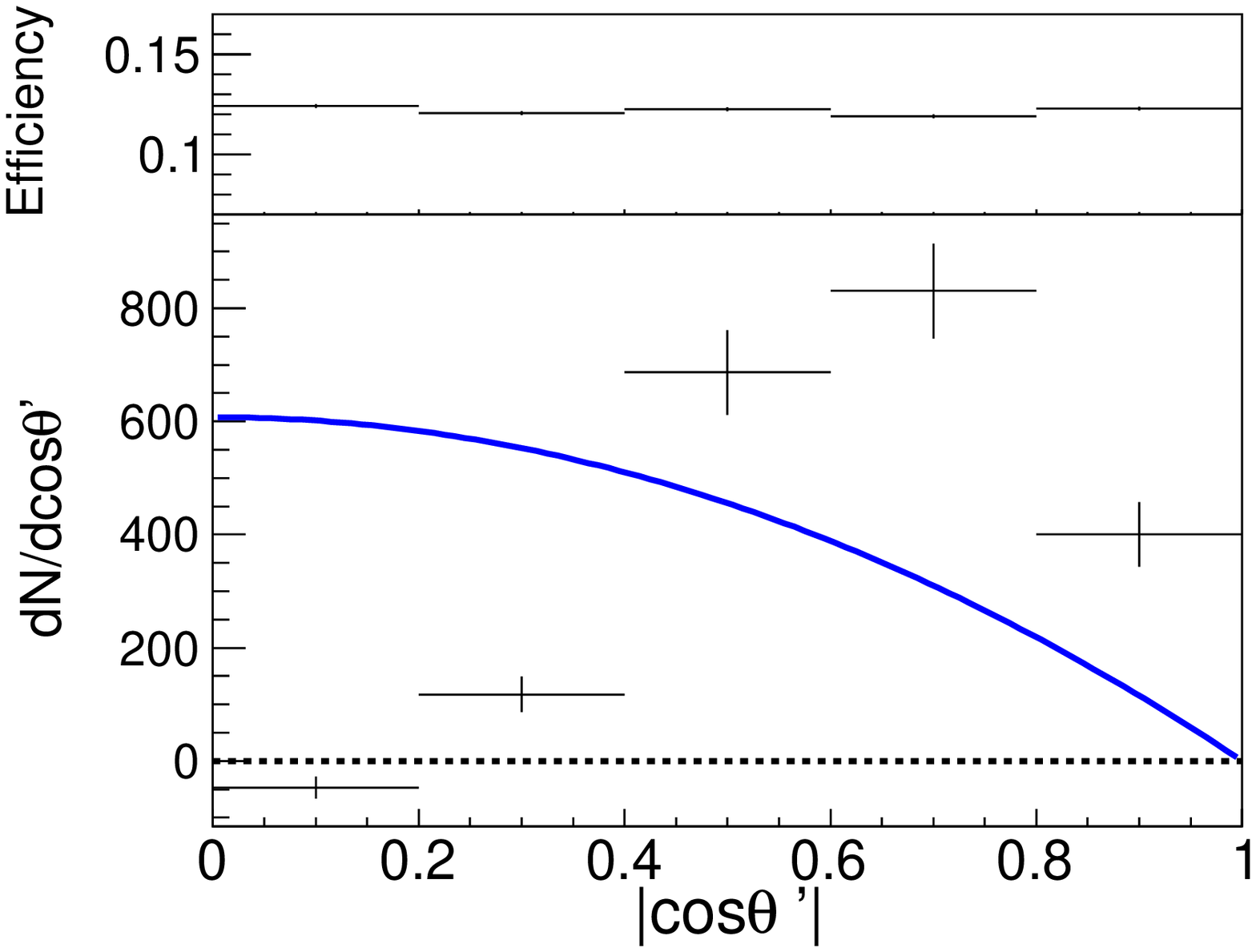}
	\put(-160, 152){\bf (a) $J^P = 1^-$}
	\put(-160, 113){\bf (b)}
	\includegraphics[width=0.48\linewidth]{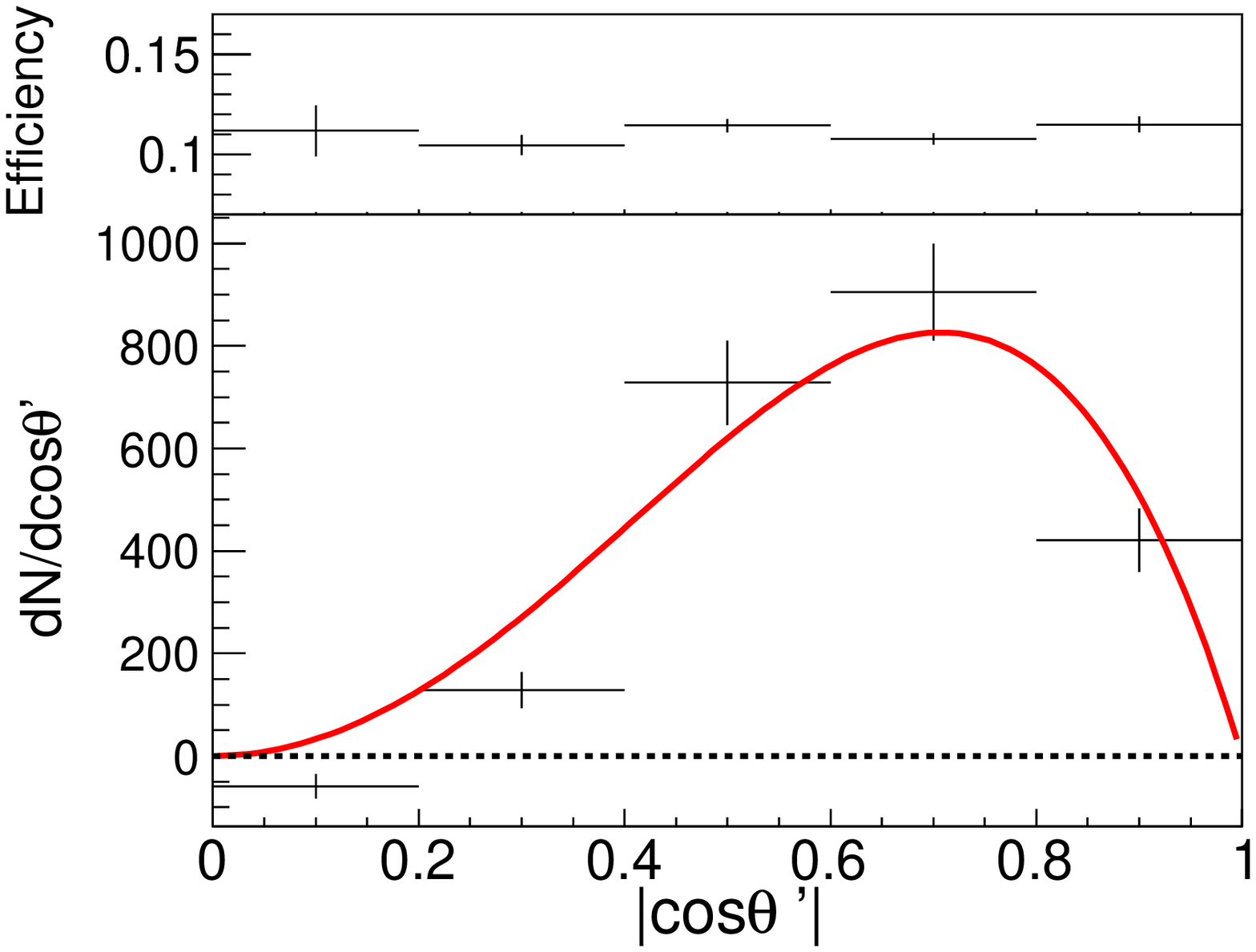}
	\put(-160, 152){\bf (c) $J^P = 2^+$}
	\put(-160, 113){\bf (d)}
	\caption{At 4.600 GeV, the efficiency-corrected $|\cos\theta'|$ distribution for the background-subtracted $\dstwom$ signals are shown in plots (b) and (d). Plots (a) and (c) are the corresponding efficiency distributions under the $J^P$ assumptions of $1^-$ and $2^+$, respectively. The shapes to be tested are shown in (b) and (d) for the two hypotheses, normalized to the area of data distribution.}
	\label{fig:jp}
\end{figure*}

In each $|\cos\theta'|$ interval of width 0.2, the number of background events is estimated from the $RQ(D_s^+)$ sideband region (2.44, 2.50) $\gevcc$ according to the global fit shown in Fig.~\ref{fig:dsfit} (d) and subtracted from the signal candidates in the signal region, (2.54, 2.60) $\gevcc$.
Then we obtain the efficiency-corrected angular distribution of $\rm d\sigma / \rm d|\cos\theta'|$, as depicted in Fig.~\ref{fig:jp} for the  $\dstwom$ signals.
The efficiency distributions in Figs.~\ref{fig:jp} (a) and  (c) are obtained from the signal MC simulation samples, which assume the spin-parity of the $\dstwom$ as $1^-$ and $2^+$, respectively.

The shapes  of the two spin-parity hypotheses are constructed as $a_{1}(1-\cos^2\theta')$ and $a_{2}\cos^2\theta'(1-\cos^2\theta')$ for $1^-$ and $2^+$, respectively.
Here, $a_{1}$ and $a_{2}$ normalize the shapes to the area of the efficiency corrected angular distributions.
To test the two different assumptions, we calculate $\chi^2 = \Sigma(\frac{y_i - \mu_i}{\sigma_i})^2$, where $i$ is the index of the interval in the angular distributions, $y_i$ is the estimated signal yield in interval $i$, $\sigma_i$ is the corresponding statistical uncertainty, and $\mu_i$ is the expected number of signal events.
The values of $\chi^2$ for the $J^P = 1^-$ and $2^+$ assumptions are evaluated as $278.67$ and $7.85$, respectively.
Hence, our results strongly favor the $J^P = 2^+$ assignment and disfavor the $J^P = 1^-$ assignment for the $\dstwom$.

\section{Studies at the other energy points}
\label{sec:crosssection}

\begin{figure*}
	\centering
	\begin{minipage}[t]{4.6cm}
	\includegraphics[height=3.cm,width=4.3cm]{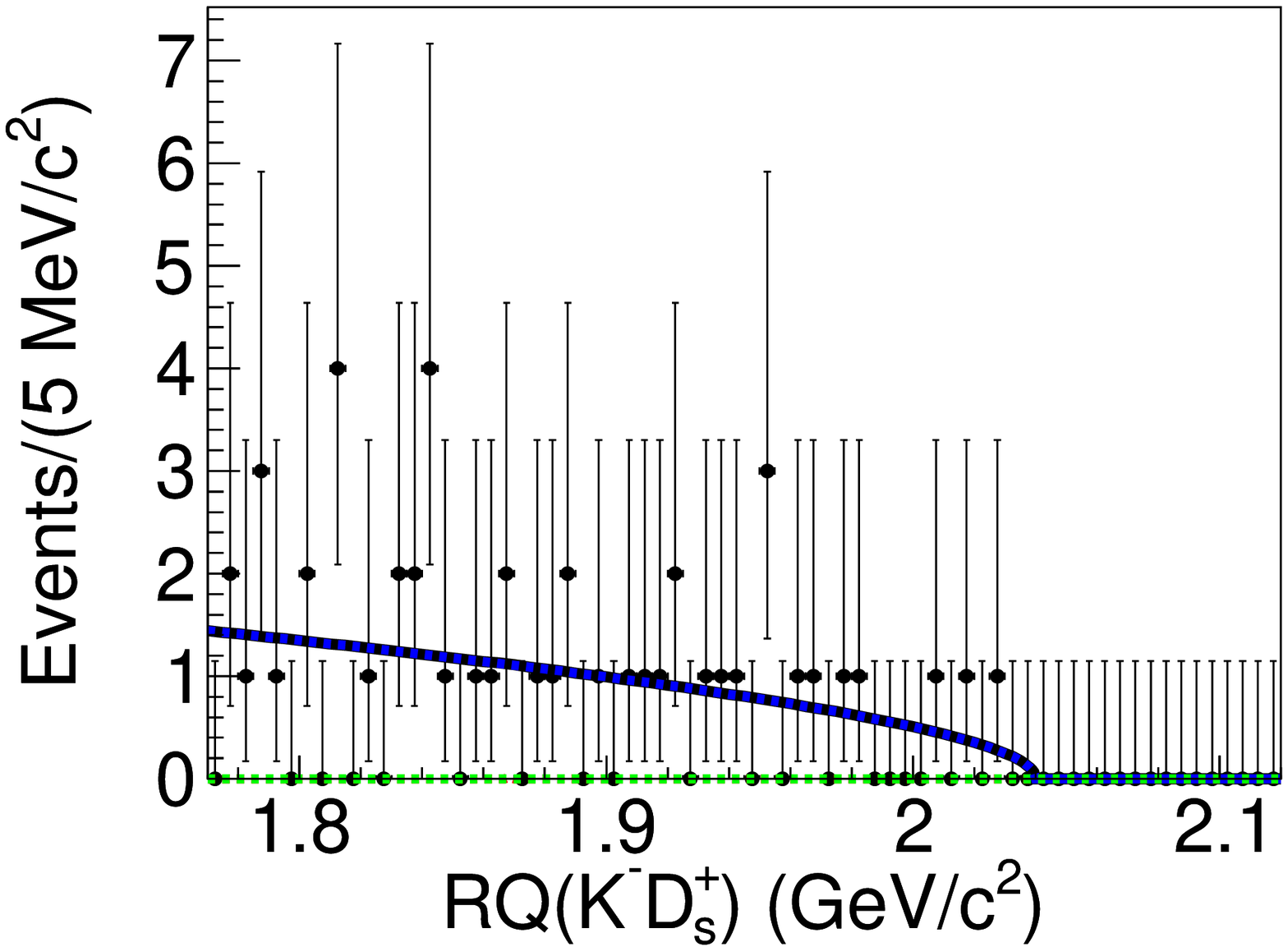}
	\put(-50, 62){4.575 GeV}
	\put(-30, 72){\bf (a)}
	\end{minipage}
	\begin{minipage}[t]{4.6cm}
	\includegraphics[height=3.cm,width=4.3cm]{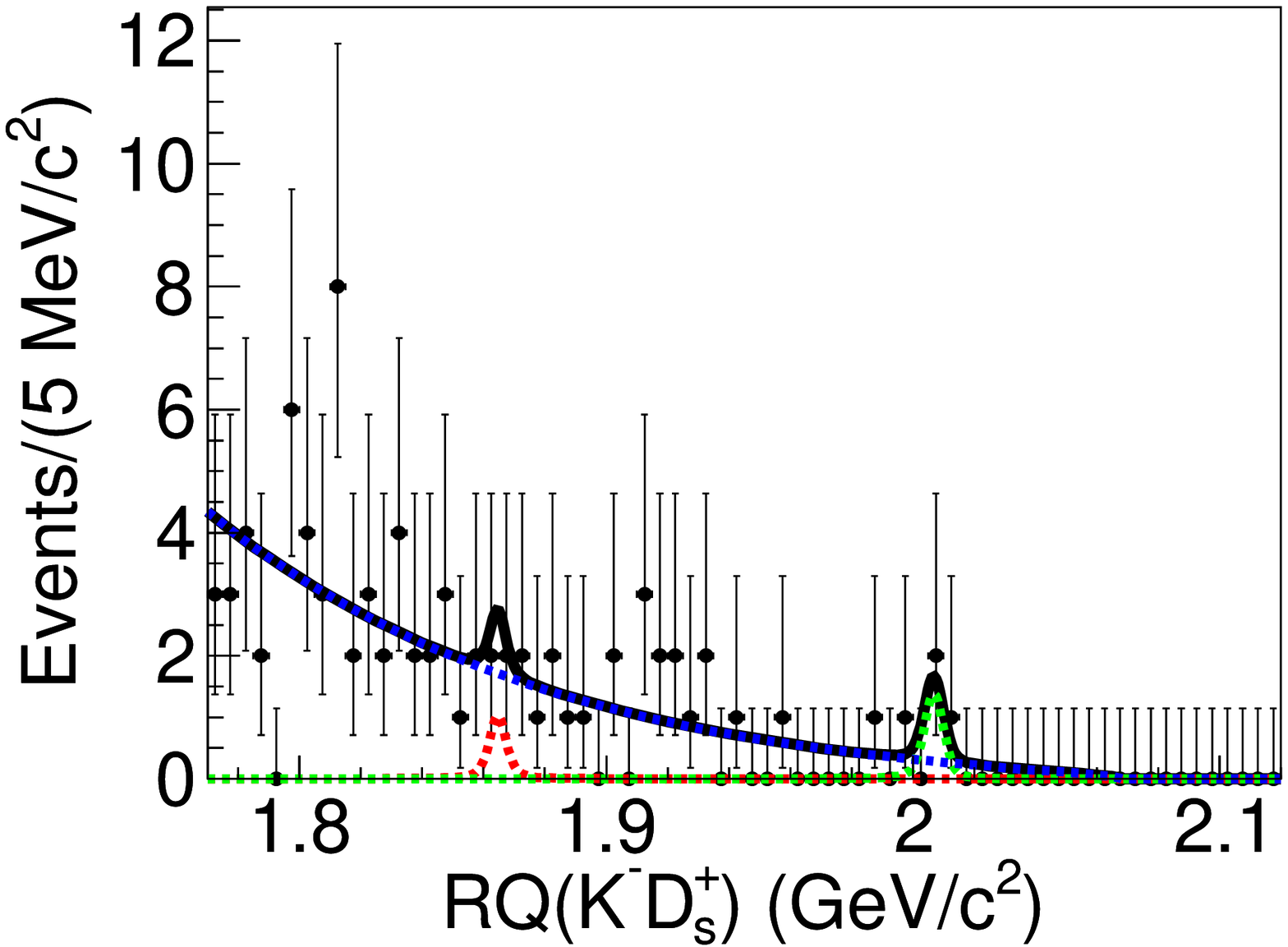}
	\put(-50, 62){4.527 GeV}
	\put(-30, 72){\bf (b)}
	\end{minipage}

	\begin{minipage}[t]{4.6cm}
	\includegraphics[height=3.cm,width=4.3cm]{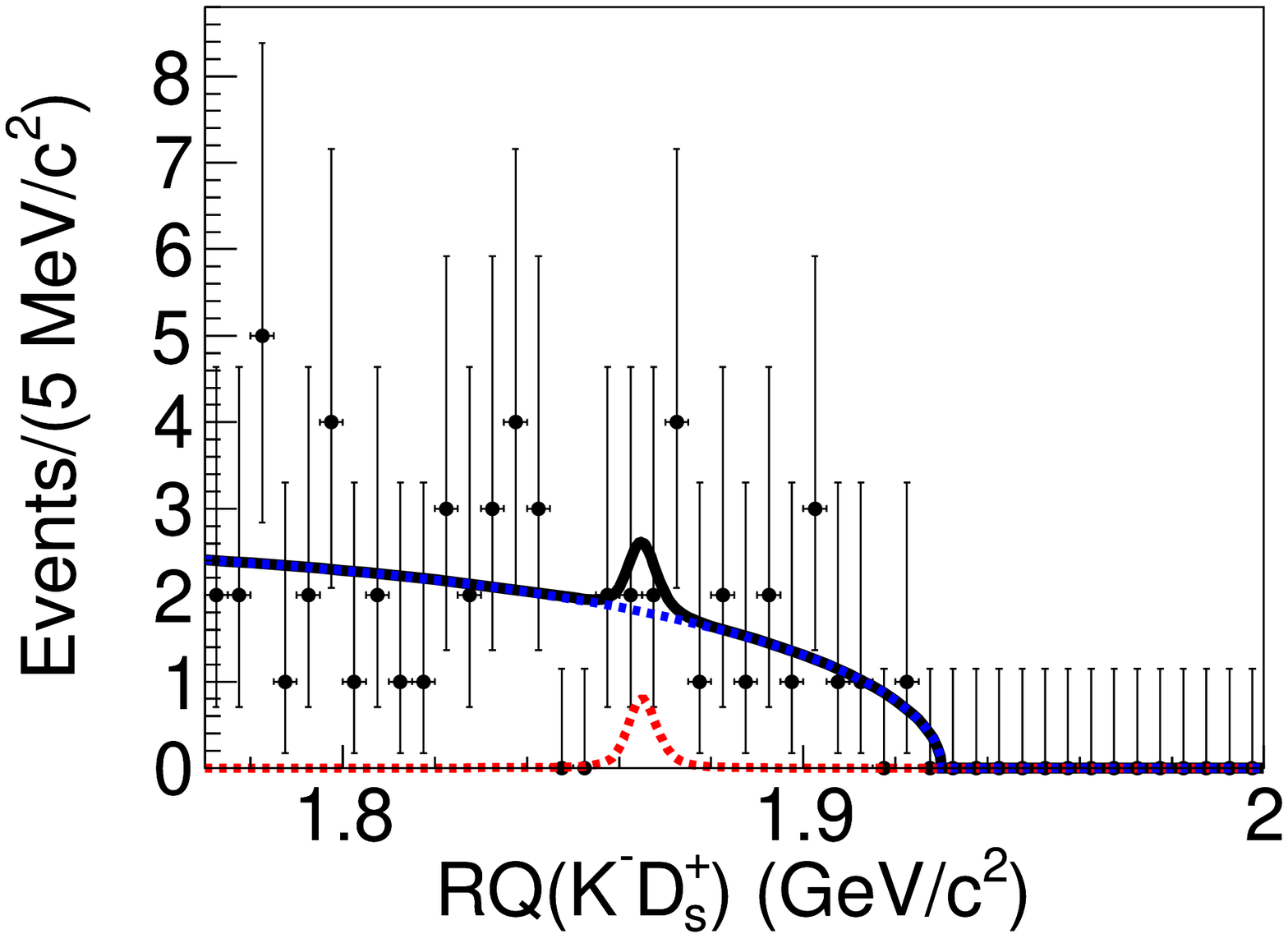}
	\put(-50, 62){4.467 GeV}
	\put(-30, 72){\bf (c)}
	\end{minipage}
	\begin{minipage}[t]{4.6cm}
	\includegraphics[height=3.cm,width=4.3cm]{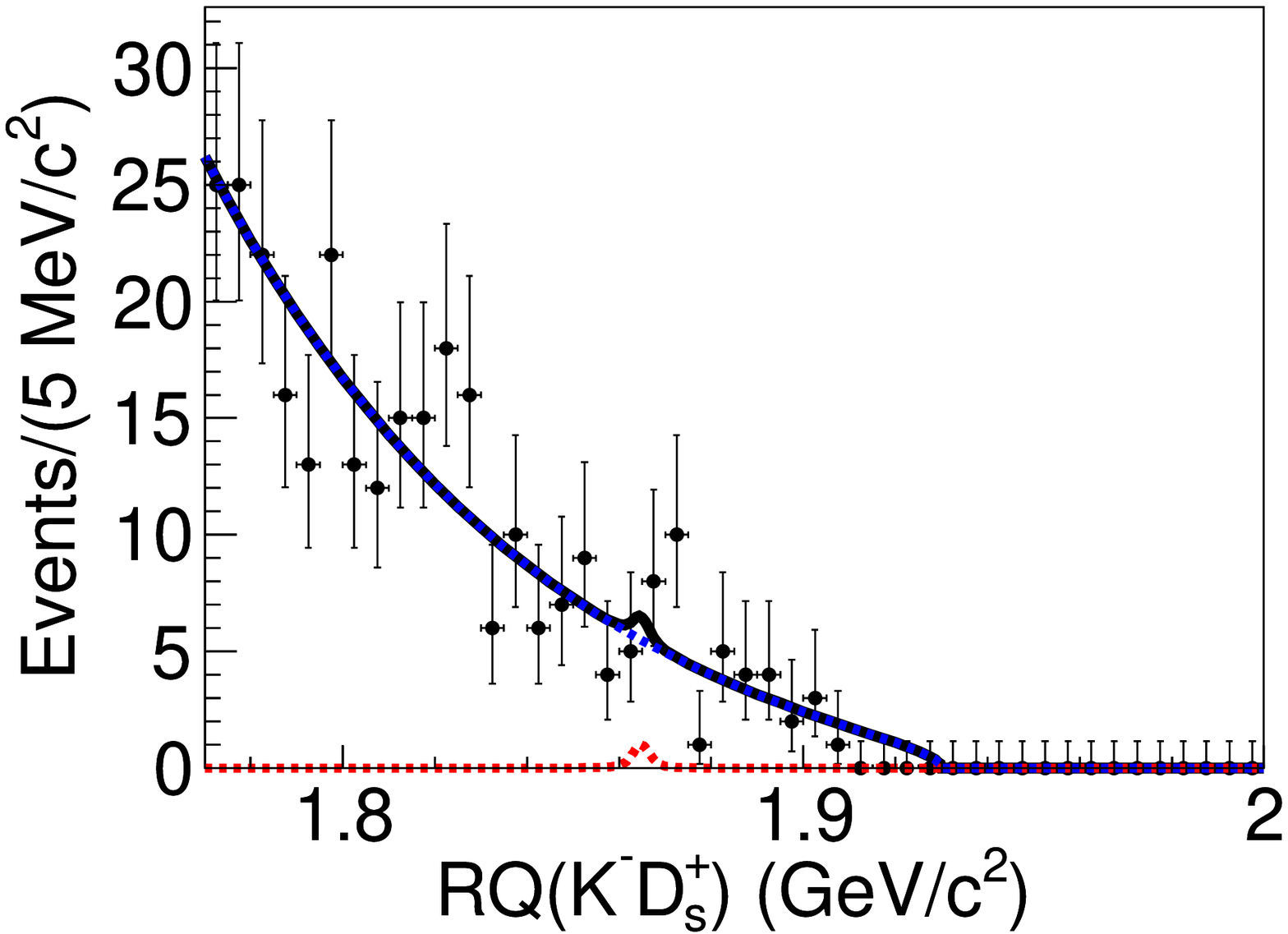}
	\put(-50, 62){4.416 GeV}
	\put(-30, 72){\bf (d)}
	\end{minipage}
	\caption{$RQ(K^- D_s^+)$ distributions and the fit results at each energy point.
Points with error bars are data, the dotted lines peaking at the nominal mass of  the $\dzerobar$($\dstzerobar$) are the signal shapes for $e^+e^-\to D_s^+ \dzerobar K^-(D_s^+ \dstzerobar K^-)$ process.}
	\label{fig:fitcrosssection}
\end{figure*}

The process $e^+e^-\to D_s^+ \overline{D}{}^{(*)0} K^-$ is also searched for at four (two) other energy points.  The corresponding integrated luminosities~\cite{lum} and center-of-mass energies~\cite{Ecms} are shown in Table~\ref{table:3bodycrosssection}.
The analysis strategy and event selection are the same as those explained in Sec.~\ref{sec:eventselection}.
The resultant $RQ(K^-D_s^+)$ distributions are shown in Fig.~\ref{fig:fitcrosssection}, together with the results of unbinned maximum likelihood fits as described in Sec.~\ref{sec:xs4600}. The fit results are given in Table~\ref{table:3bodycrosssection}.

As has been studied with the largest statistics data at $\sqrt{s}=4.600\gev$, the processes $D_s^+\dsonem$ and $D_s^+\dstwom$ dominate the processes $e^+e^-\to D_s^+ \overline{D}{}^{*0} K^-$ and $e^+e^-\to D_s^+ \overline{D}{}^{0} K^-$, respectively.
We assume that this conclusion still holds for the MC simulations of the final states of $D_s^+ \overline{D}{}^{(*)0} K^-$ for the energy points above the $D_s^+\dsonem$ or $D_s^+\dstwom$ mass thresholds.
For the energy points below the mass thresholds, the signal MC simulation samples of the three-body processes are generated with average momentum distributions in the phase space.

Since the four data samples taken at lower energies suffer from low statistics,
we also present upper limits at the $90\%$ confidence level on the cross sections.
The upper limits are determined using a Bayesian approach with a flat prior.
The systematic uncertainties are considered by convolving the likelihood distribution with a Gaussian function representing the systematic uncertainties.
The numerical results are summarized in Table~\ref{table:3bodycrosssection}.

\begin{table*}
\centering
\caption{Cross section measurements at different energy points. For the cross sections, the first set of uncertainties are statistical and the second are systematic. The uncertainties of the number of observed signals are statistical only. The four samples with lower center-of-mass energies suffer from low statistics, we therefore set the lower and upper boundary of the uncertainties of $N_{\rm obs}$ as 0 and the upper limits at the 68.3\% confidence level, respectively. }\label{table:3bodycrosssection}
\begin{tabular}{c|l|rrrrr}
\hline\hline
   & $\sqrt{s}~(\gev)$               &  4.600            & 4.575                       & 4.527                         & 4.467                        & 4.416 \\
\hline
   &  $\mathcal{L}$~($\rm pb^{-1}$)  &  567           & 48                       & 110                        & 110                       & 1029   \\
   &  $\frac{1}{|1-\Pi|^2}$         &  1.059            & 1.059                       & 1.059                         & 1.061                        & 1.055   \\
\hline
   &  $1+\delta$                    &  0.765            & 0.755                       & 0.735                         &                              &     \\
	&  $\epsilon$(\%)               &  16.1             & 14.3                        & 13.2                          &                              &    \\
$\dspp \dstzerobar K^-$   & $N_{\rm obs}$& $41.0\pm9.3$     & $0.0_{-0.0}^{+2.0}$         & $2.3_{-2.3}^{+3.9}$           &                              &   \\
   &  $\sigma^B$~($\rm pb$)          &$10.1\pm2.3\pm0.8$ & $0.0_{-0.0-0.0}^{+7.3+1.1}$ & $3.9_{-3.9}^{+6.6}\pm 0.4$    &                              &   \\
   &  $N^{\rm up}$                      &                   &  3.7                        &  6.7                          &                              &   \\
	&  $\sigma^{B}_{U.L.}~($\rm pb$)$ &                  &  13.5                       &  11.3                         &                              &   \\
\hline
   &  $1+\delta$                    & 0.694             & 0.698                       &  0.702                        &  0.691                       & 0.762 \\
	&  $\epsilon$(\%)               & 22.3              & 23.9                        &  20.3                         &  18.2                        & 14.6 \\
$\dspp \dzerobar K^-$ & $N_{\rm obs}$   & $98.4\pm 11.7$    & $0.0_{-0.0}^{+3.0}$         & $1.7_{-1.7}^{+4.5}$           &  $4.1_{-4.1}^{+7.1}$         & $1.2_{-1.2}^{+8.0}$   \\
   &  $\sigma^B$~($\rm pb$)          & $19.4\pm2.3\pm1.6$& $0.0_{-0.0-0.0}^{+6.5+0.9}$ & $1.9_{-1.9}^{+5.0}\pm 0.2$    &  $5.1_{-5.1}^{+8.9} \pm 0.4$ & $0.3_{-0.3}^{+1.2}\pm 0.1$  \\
   &  $N^{\rm up}$                      &                   &  5.8                        &  7.3                          &  10.6       		         & 10.5   \\
   &  $\sigma^{B}_{U.L.}~($\rm pb$)$ &                   &  12.7                       &  8.1                          &  13.2                        & 1.6    \\

 \hline\hline
\end{tabular}
\end{table*}


\begin{table*}
	\centering
	\caption{Summary of systematic uncertainties on the $\dsonem$ and $\dstwom$ resonance parameters measured at $\sqrt{s}=4.600\gev$. ``$\cdots$'' means the uncertainty is negligible.}\label{table:ds1ds2uncertainty}
	\begin{tabular}{l|cc|cc}
		\hline\hline
		 & \multicolumn{2}{c|}{Mass~($\rm\mevcc$)   } & \multicolumn{2}{c}{Width~(\mev)}\\
		\hline
		 Source                    &  $\dsonem$ & $\dstwom$   &  $\dsonem$ & $\dstwom$ \\
		 \hline
		 Mass shift  &    3.0   &     1.3    &  $\cdots$ & $\cdots$  \\
		 Detector resolution        &$\cdots$  & $\cdots$   &    0.5    &  0.1      \\
                 Center-of-mass energy      &  0.7     &  1.0       &    0.2    &  0.3      \\
		 Signal model               &$\cdots$  &            & $\cdots$  &           \\
		 Background shape           &    0.2   &     0.4    &    0.2    &  0.3      \\
		 Fit range                  &$\cdots$  & $\cdots$   &    0.2    &  1.0      \\
		 \hline
                 Total                      &    3.1   &    1.7     &    0.6    &   1.1     \\
		 \hline\hline
	\end{tabular}
\end{table*}

\begin{table*}
	\centering
	\caption{Relative systematic uncertainties (in \%) on the cross section measurement. The first value in brackets is for $\dspp\dzerobar K^-$, and the second  for $\dspp\dstzerobar K^-$. ``$\cdots$'' means the uncertainty is negligible. ``-'' means unavailable due to $\sqrt{s}$ being below the production threshold.}
	\label{table:3bodycrosssectionErr}
	\begin{tabular}{l|c|c|c|c|c||c|c}
		\hline
		\hline
		 & \multicolumn{5}{c||}{$\sigma^B(e^+e^- \to \dspp \overline{D}^{(*)0} K^-)$ at different $\sqrt{s} (\gev)$} & \multicolumn{2}{c}{$e^+e^-\to D_s^+D_{sJ}^-$ at 4.600 GeV}\\
		 \hline
		 Source      &   4.600  &  4.575  & 4.527 & 4.467 & 4.416 & $\dsonem$ & $\dstwom$\\
		 \hline
		 Tracking    &   4     &   4    &  4 &  4 &  4  & 4 & 4 \\
		 Particle ID &   4     &   4    &  4 &  4 &  4  & 4 & 4 \\
		 Luminosity  &   1     &   1    &  1 &  1 &  1  & 1 & 1  \\
		 Branching faction&  3 &   3    &  3 &  3 &  3  & 3 & 3  \\
		 center-of-mass energy& $\cdots$& $\cdots$& $\cdots$& $\cdots$& $\cdots$& $\cdots$ & $\cdots$\\
		 Fit range   &   ($\cdots$, 2)   & (2, $\cdots$) & (4, 3)   &  ($\cdots$,-)   & ($\cdots$,-)  & 3 & 4 \\
		 Background shape & (3, 1)   &  (1, 4)      & (4, 5)   & (5,-)    & (6,-) &   4  & 5  \\
		 Line shape &  (3, 4)  &  (2, 3)      &  (1, 1)  & (1,-)   & ($\cdots$,-)  &  4   & 3 \\
		 \hline
		 Total: &     (8, 8)           &  (7, 8)      & (9, 9)   & (8,-)   & (9,-) &  9  & 10 \\
		 \hline\hline
	\end{tabular}
\end{table*}

\section{Systematic Uncertainties}
The systematic uncertainties on the resonance parameters and cross section measurements are summarized in Tables~\ref{table:ds1ds2uncertainty} and \ref{table:3bodycrosssectionErr}, respectively, where the total systematic uncertainties are obtained by adding all items in quadrature. For each item, details are elaborated as follows.

\begin{enumerate}
	\item \emph{Tracking efficiency.}  The difference in tracking efficiency for the kaon and pion reconstruction between the MC simulation and the real data is estimated to be $1.0\%$ per track~\cite{trkpiderr}. Hence, $4.0\%$ is taken as the systematic uncertainty for four charged tracks.
	\item \emph{PID efficiency.}  The uncertainty of identifying the particle types of kaon and pion is estimated to be $1\%$ per charged track~\cite{trkpiderr}. Therefore, $4.0\%$ is taken as the systematic uncertainty for the PID efficiency of the four detected charged tracks.
	\item \emph{Signal Model.}  In the fits of the $\dsonem$, the fraction of the $D$-wave and $S$-wave components is varied according to the Belle measurement~\cite{Ds1mixing}, and the maximum changes on the fit results are taken as systematic uncertainties.
In the measurement of the $\dstwom$ resonance parameters, the uncertainty stemming from the signal model is negligible as the $D$-wave amplitude dominates in the heavy quark limit.
	\item \emph{Background Shape.} In the measurements of the $\dsonem$ and $\dstwom$ resonance parameters,
linear background functions are used in the nominal fits.
To estimate the uncertainties due to the background parametrization,
higher order polynomial functions are studied, and the largest changes on the final results are taken as the systematic uncertainty.
In the measurement of $\sigma^B(e^+e^-\to D_s^+ \overline{D}{}^{(*)0} K^-$),
we replace the ARGUS background shape in the nominal fit with a second-order polynomial function $a(m-m_0)^2 + b$, where $m_0$ is the threshold value and is the same as that in the nominal fit, while $a$ and $b$ are free parameters.
We take the difference on the final results as the systematic uncertainty.
	\item \emph{Fit Range.} We vary the boundaries of the fit ranges to estimate the relevant systematic uncertainty, which are taken as the maximum changes on the numerical results.
	\item \emph{Mass Shift and Detector Resolution.} In the nominal fits to measure the $\dsonem$ and $\dstwom$ resonance parameters, the effects of a mass shift and the detector resolution are included in the MC determined detector resolution shape. The potential bias from the MC simulations are studied using the control sample of $e^+e^-\to D_s^+ D_s^{*-}$. We select the $D_s^+$ candidates following the aforementioned selection criteria and plot the $RQ(D_s^+)$ distribution to be fitted to the $D_s^{*-}$ peak. The signal function is composed of a Breit-Wigner shape convolved with a Gaussian function.
We extract the detector resolution parameters from a series of fits at different momentum intervals of the $D_{s}^+$ candidates. Hence, the absolute  resolution parameters for the fits to the $\dsonem$ or $\dstwom$ are extrapolated according to the detected $D_{s}^+$ momentum.
In an alternative fit, we fix the resolution parameters according to this study, instead of to the MC-determined resolution shape.
The resultant change in the new fit from the original fit is considered as the systematic uncertainty.
	\item \emph{Branching Fraction.} The systematic uncertainty in the branching fraction for the process $D_s^+ \to K^+ K^-\pi^+$ is taken from PDG~\cite{pdg}.
	\item \emph{Luminosity.} The integrated luminosity of each sample is measured with a precision of $1\%$ with Bhabha scattering events~\cite{lum}.
	\item \emph{Center-of-mass energy.} We change the values of center-of-mass energy of each sample according to the uncertainties in Ref.~\cite{Ecms} to estimate the systematic uncertainties due to the center-of-mass energy.
	\item \emph{Line Shape of Cross Section.} The line shape of the $e^+e^-\to D_s^+ \overline{D}{}^{(*)0}K^-$ cross section  (including the intermediate $\dsonem$ and $\dstwom$ states) affects the radiative correction factor and the detection efficiency.
This uncertainty is estimated by changing the input of the observed line shape to the simulation.
In the nominal measurement, a power function of $c\cdot(\sqrt{s}-E_0)^d$ is taken as the input of the observed line shape.
Here, $E_0$ is the production threshold energy for the process $e^+e^-\to D_s^+  \overline{D}{}^{(*)0} K^-$, and $c$ and $d$ are parameters determined from fits to the observed line shape.
To estimate the uncertainty, we change the exponent of the nominal input power function to $d\pm1$ and compare the results with the nominal measurement. The largest difference is taken as the systematic uncertainty.
\end{enumerate}

\section{Summary}

We study the process $e^+e^-\to \dspp \overline{D}{}^{(*)0} K^-$ at 4.600 GeV and observe the two $P$-wave charmed-strange mesons,
$\dsonem$ and $\dstwom$.
The $\dsonem$ mass is measured to be $(2537.7 \pm 0.5 \pm 3.1)~\rm\mevcc$ and its width is $(1.7\pm 1.2 \pm 0.6)~\rm\mev$, both consistent with the current world-average values in PDG~\cite{pdg}. The mass and width of the $\dstwom$ meson are measured to be $(2570.7\pm 2.0 \pm 1.7)~\rm\mevcc$ and $(17.2 \pm 3.6 \pm 1.1)~\rm\mev$, respectively, which are compatible with the LHCb~\cite{lhcds2mw,Aaij:2014xza} and PDG~\cite{pdg} values.
The spin-parity of the $\dstwom$ meson is determined to be $J^P=2^{+}$, which confirms the LHCb result~\cite{lhcbds2jp}.
The Born cross sections are measured to be $\sigma^{B}(e^+e^-\to\dspp\dstzerobar K^-) = (10.1\pm2.3\pm0.8)~\rm\pb$ and $\sigma^{B}(e^+e^-\to\dspp \overline{D}{}^{0} K^-) = (19.4\pm2.3\pm1.6)~\rm\pb$.
The products of the Born cross section and the decay branching fraction are measured to be
$\sigma^{B}(e^+e^-\to\dspp\dsonem + c.c.)\cdot\mathcal{B}(\dsonem\to\dstzerobar K^-) = (7.5 \pm 1.8 \pm 0.7)~\rm\pb$
and
$\sigma^{B}(e^+e^-\to\dspp\dstwom + c.c.)\cdot\mathcal{B}(\dstwom\to\dzerobar K^-) = (19.7 \pm 2.9 \pm 2.0)~\rm\pb$.
In addition, the processes $e^+e^-\to \dspp \overline{D}{}^{(*)0} K^-$ are searched for using small data samples taken at four (two) center-of-mass energies between 4.416 (4.527) and 4.575 GeV, and upper limits at the $90\%$ confidence level on the cross sections are determined.

\section{Acknowledgments}

The BESIII collaboration thanks the staff of BEPCII and the IHEP computing center for their strong support. This work is supported in part by National Key Basic Research Program of China under Contract No. 2015CB856700; National Natural Science Foundation of China (NSFC) under Contracts Nos. 11335008, 11425524, 11625523, 11635010, 11735014; the Chinese Academy of Sciences (CAS) Large-Scale Scientific Facility Program; the CAS Center for Excellence in Particle Physics (CCEPP); Joint Large-Scale Scientific Facility Funds of the NSFC and CAS under Contracts Nos. U1532257, U1532258, U1732263; CAS Key Research Program of Frontier Sciences under Contracts Nos. QYZDJ-SSW-SLH003, QYZDJ-SSW-SLH040; 100 Talents Program of CAS; INPAC and Shanghai Key Laboratory for Particle Physics and Cosmology; German Research Foundation DFG under Contracts Nos. Collaborative Research Center CRC 1044, FOR 2359; Istituto Nazionale di Fisica Nucleare, Italy; Koninklijke Nederlandse Akademie van Wetenschappen (KNAW) under Contract No. 530-4CDP03; Ministry of Development of Turkey under Contract No. DPT2006K-120470; National Science and Technology fund; The Swedish Research Council; U. S. Department of Energy under Contracts Nos. DE-FG02-05ER41374, DE-SC-0010118, DE-SC-0010504, DE-SC-0012069; University of Groningen (RuG) and the Helmholtzzentrum fuer Schwerionenforschung GmbH (GSI), Darmstadt.

\end{multicols}


\vspace{10mm}
\vspace{-1mm}
\centerline{\rule{80mm}{0.1pt}}
\vspace{2mm}

\begin{multicols}{2}

\end{multicols}
\end{CJK*}


\begin{thebibliography}{99}


\bibitem{model1} S.~Godfrey and N.~Isgur, Phys.\ Rev.\ D {\bf 32}, 189 (1985).
\bibitem{model2} N.~Isgur and M.~B.~Wise, Phys.\ Rev.\ Lett.\  {\bf 66}, 1130 (1991).
\bibitem{model3} J.~L.~Rosner, Comments Nucl.\ Part.\ Phys.\  {\bf 16}, 109 (1986).
\bibitem{model4} M.~Di Pierro and E.~Eichten, Phys.\ Rev.\ D {\bf 64}, 114004 (2001).

\bibitem{pdg} M.~Tanabashi {\it et al.} (Particle Data Group), Phys.\ Rev.\ D {\bf 98}, 010001 (2018).

\bibitem{lhcds2mw} R.~Aaij {\it et al.} (LHCb Collaboration), Phys. \ Lett. \ B {\bf 698}, 14 (2011).
\bibitem{Aaij:2014xza} R.~Aaij {\it et al.} (LHCb Collaboration), Phys.\ Rev.\ Lett.\  {\bf 113}, 162001 (2014).

\bibitem{pdgDiff1} B.~Aubert {\it et al.} (BABAR Collaboration), Phys.\ Rev.\ Lett.\ {\bf 97}, 222001 (2006).
\bibitem{pdgDiff2} H.~Albrecht {\it et al.} (ARGUS collaboration), Z.\ Phys.\ C {\bf 69}, 405 (1996).
\bibitem{pdgDiff3} Y.~Kubota {\it et al.} (CLEO collaboration), Phys.\ Rev.\ Lett.\ {\bf 72}, 1972 (1994).

\bibitem{lhcbds2jp} R.~Aaij {\it et al.} (LHCb Collaboration), Phys. \ Rev. \ D. {\bf 90}, 072003 (2014).

\bibitem{besRvalue} M.~Ablikim {\it et al.} (BESIII Collaboration), Phys.\ Lett. \ B {\bf 660}, 315 (2008)
\bibitem{Y1} B.~Aubert  {\it et al.} (BABAR Collaboration),  Phys.\ Rev.\ Lett.\ {\bf 95}, 142001 (2005).
\bibitem{Y2} T.E.~Coan {\it et al.}  (CLEO Collaboration), Phys.\ Rev.\ Lett.\ {\bf 96}, 162003 (2006).
\bibitem{Y3} C.Z.~Yuan {\it et al.}  (Belle Collaboration), Phys.\ Rev.\ Lett.\ {\bf 99}, 182004 (2007).
\bibitem{Y4} B.~Aubert {\it et al.} (BABAR Collaboration), Phys.\ Rev.\ Lett.\ {\bf 98},212001 (2007).
\bibitem{Y5} X.L.~Wang {\it et al.} (Belle Collaboration), Phys.\ Rev.\ Lett.\ {\bf 99}, 142002 (2007).

\bibitem{dd1} G.~Pakhlova {\it et al.} (Belle Collaboration), Phys.\ Rev.\ D {\bf 77}, 011103 (2008).
\bibitem{dd2} G.~Pakhlova {\it et al.} (Belle Collaboration), Phys.\ Rev.\ Lett.\ {\bf 98}, 092001 (2007).
\bibitem{dd3} G.~Pakhlova {\it et al.} (Belle Collaboration), Phys.\ Rev.\ Lett.\ {\bf 100}, 062001 (2008).
\bibitem{dd4} G.~Pakhlova {\it et al.} (Belle Collaboration), Phys.\ Rev.\ D {\bf 80}, 091101 (2009).
\bibitem{dd5} G.~Pakhlova {\it et al.} (Belle Collaboration), Phys.\ Rev.\ Lett.\ {\bf 101}, 172001 (2008).
\bibitem{dd6} G.~Pakhlova {\it et al.} (Belle Collaboration), Phys.\ Rev.\ D {\bf 83}, 011101 (2011).
\bibitem{dd7} B.~Aubert {\it et al.} (BABAR Collaboration), Phys.\ Rev.\ D {\bf 76}, 111105 (2007).
\bibitem{dd8} B.~Aubert {\it et al.} (BABAR Collaboration), Phys.\ Rev.\ D {\bf 79}, 092001 (2009).
\bibitem{dd9} P.~del~Amo~Sanchez {\it et al.} (BABAR Collaboration), Phys.\ Rev.\ D {\bf 82}, 052004 (2010).
\bibitem{dd10} D.~Cronin-Hennessy {\it et al.} (CLEO Collaboration), Phys.\ Rev.\ D {\bf 80}, 072001 (2009).
\bibitem{dd11} M.~Ablikim {\it et al.} (BESIII Collaboration), Phys.\ Lett. \ Lett. {\bf 112}, 022001 (2014).
\bibitem{dd12} M.~Ablikim {\it et al.} (BESIII Collaboration), Phys.\ Lett. \ Lett. {\bf 115}, 222002 (2015).
\bibitem{dd13} M.~Ablikim {\it et al.} (BESIII Collaboration), Phys.\ Lett. \ D {\bf 92}, 092006 (2015).
\bibitem{dd14} M.~Ablikim {\it et al.} (BESIII Collaboration), arXiv:1808.02847.
\bibitem{d2d} G.~Pakhlova {\it et al.} (Belle Collaboration), Phys.\ Rev.\ Lett.\ {\bf 100}, 062001 (2008).

\bibitem{lum} M.~Ablikim {\it et al.} (BESIII Collaboration), Chin.\ Phys.\ C {\bf 39}, 093001 (2015).
\bibitem{Ecms} M.~Ablikim {\it et al.} (BESIII Collaboration), Chin.\ Phys.\ C {\bf 40}, 063001 (2016).
\bibitem{Ablikim:2009aa} M.~Ablikim {\it et al.} (BESIII Collaboration), Nucl.\ Instrum.\ Meth.\ A {\bf 614}, 345 (2010).
\bibitem{Yu:IPAC2016-TUYA01} C.~H.~Yu {\it et al.} Proceedings of IPAC2016, Busan, Korea, 2016, doi:10.18429/JACoW-IPAC2016-TUYA01.
\bibitem{geant4} S.~Agostinelli {\it et al.} (GEANT4 Collaboration), Nucl.\ Instrum.\ Meth.\ A {\bf 506}, 250 (2003).
\bibitem{ref:kkmc} S.~Jadach, B.~F.~L.~Ward, and Z.~Was, Comput.\ Phys.\ Commun. {\bf 130}, 260 (2000); Phys.\ Rev.\ D {\bf 63}, 113009 (2001).
\bibitem{ref:evtgen} D.~J.~Lange, Nucl.\ Instrum.\ Meth.\ A {\bf 462}, 152 (2001); R.~G.~Ping, Chin. Phys. C {\bf 32}, 599 (2008).
\bibitem{ref:lundcharm} J.~C.~Chen, G.~S.~Huang, X.~R.~Qi, D.~H.~Zhang and Y.~S.~Zhu, Phys.\ Rev.\ D {\bf 62}, 034003 (2000); R.~L.~Yang, R.~G.~Ping and H.~Chen, Chin.\ Phys.\ Lett.\  {\bf 31}, 061301 (2014).
\bibitem{photos} E.~Richter-Was, Phys.\ Lett.\ B {\bf 303}, 163 (1993).
\bibitem{dsdalitz} R.~E.~Mitchell {\it et al.} (CLEO Collaboration), Phys.\ Rev.\ D {\bf 79}, 072008 (2009).
\bibitem{hqrev} N.~Brambilla {\it et al.},  Eur.\ Phys.\ J.\ C  {\bf 71}, 1534 (2011).
\bibitem{lcbf} M.~Ablikm {\it et al.} (BESIII Collaboration),  Phys.\ Rev.\ Lett.\ {\bf 116}, 052001 (2016).
\bibitem{nima462.152} D.~J.~Lange, Nucl.\ Instrum.\ Meth.\ A {\bf 462}, 152 (2001); R.~G.~Ping, Chin.\ Phys.\ C {\bf 32}, 599 (2008).
\bibitem{Ds1mixing}	V.~Balagura {\it et al.} (Belle Collaboration), Phys.\ Rev.\ D  {\bf 77}, 032001 (2008).
%
\bibitem{Kuraev:1985hb} E.~A.~Kuraev and V.~S.~Fadin, Sov.\ J.\ Nucl.\ Phys.\  {\bf 41}, 466 (1985) [Yad.\ Fiz.\  {\bf 41}, 733 (1985)].
\bibitem{vacuum.polarization} F.~Jegerlehner, Z.\ Phys.\ C  {\bf 32}, 195 (1986).
\bibitem{trkpiderr} M.~Ablikim {\it et al.} (BESIII Collaboration), Phys.\ Rev.\ Lett.\ {\bf 112}, 022001 (2014).

\bibitem{argus} H.~Albrecht {\it et al.} (ARGUS Collaboration), Phys.\ Lett.\ B {\bf 340}, 217 (1994).

\end{thebibliography}
\end{document}